\documentclass[sn-mathphys,iicol]{sn-jnl}

\jyear{2021}%

\begin{document}

\title[\hspace{4in} Orbital antiferroelectricity and higher dimensional magnetoeletric order in a spin chain]{Orbital antiferroelectricity and higher dimensional magnetoelectric order in the spin-$1/2$ XX chain extended with three-spin interactions}

\author[1,2]{\fnm{Pradeep} \sur{Thakur}}\email{pskvkthakur@gmail.com}
\equalcont{These authors contributed equally to this work.}

\author*[1]{\fnm{Durganandini} \sur{P}}\email{pdn@physics.unipune.ac.in}
\equalcont{These authors contributed equally to this work.}

\affil*[1]{\orgdiv{Department of Physics}, \orgname{Savitribai Phule Pune University}, \orgaddress{\street{Ganeshkhind Road}, \city{Pune}, \postcode{411007}, \state{Maharashtra}, \country{India}}}

\affil[2]{\orgdiv{Department of Physics}, \orgname{COEP Technological University}, \orgaddress{\street{Wellesley Road}, \city{Shivajinagar, Pune}, \postcode{411005}, \state{Maharashtra}, \country{India}}}

%%===============================%%
%% Aliases for long instructions %%
%%===============================%%

\newcommand{\be}{\begin{equation}}
\newcommand{\ee}{\end{equation}}
\newcommand{\bea}{\begin{eqnarray}}
\newcommand{\eea}{\end{eqnarray}}
\newcommand{\del}{\partial}
\newcommand{\lang}{\langle}
\newcommand{\rang}{\rangle}
\newcommand{\im}{\imath}
\newcommand{\e}{\epsilon}
\newcommand{\om}{\omega}
\newcommand{\s}{\sigma}
\newcommand{\ek}{\epsilon _k}

\abstract{We study the spin-1/2 XX model extended with three-spin interactions of the XZX+YZY and XZY-YZX types. We solve the model exactly and obtain the ground state phase diagram as a function of the two three-spin coupling strengths. 
We show that even in absence of external electric and magnetic fields there is a phase which exhibits spontaneous magnetoelectric order when both XZX+YZY and XZY-YZX interactions are present. Specifically, in this regime, we show that there exists not only a non-zero magnetization and a scalar chirality but also a vector chiral order. Further, we show the existence of a plaquette vector chirality, or circulating chiral spin current loops, in the plaquettes n, n+1, n+2 with the sense of the current being opposite in adjacent plaquettes. Analogous to charge current loops giving rise to orbital magnetic dipole moments, the circulating spin current loops give rise to orbital electric dipole moments - a novel orbital antiferroelectricity. We characterize this phase by a higher-dimensional scalar and vector toroidal order. Such a novel phase with higher dimensional order arises because of the non-trivial topological connectivity resulting from the presence of both the three-spin interactions. We also study the combined effect of both types of three-spin interactions on the entanglement entropy.}


\maketitle

\section{Introduction}\label{sec1}

In recent years, the interplay of electric and magnetic order in quantum spin systems has attracted a lot of attention, motivated by the direct correlation between ferroelectricity and noncollinear spin order discovered in several multiferroics~\cite{mf-review}. 
This direct correlation implies the existence of an intrinsic magnetoelectric effect (MEE) in these systems~\cite{curieMEE,fiebig2005}.
Studies of the MEE are of great experimental interest today due to the possibility of being able to manipulate magnetism with electric fields~\cite{fina2017}, which is very useful for practical applications such as in magnetic storage devices, magnetic memories~\cite{noda2007} etc.

The spin-$1/2$ Heisenberg Hamiltonian, with bilinear interactions between nearest-neighbouring spins, has occupied a paradigmatic position in theoretical studies of frustrated magnetism~\cite{mikeska-review}.
When spin-orbit interactions are taken into account at the atomic level, an additional spin exchange interaction emerges in the effective spin Hamiltonian, the Dzyaloshinskii-Moriya interaction (DMI)~\cite{DM1,DM2}:
 $$
 H_{DM}= \vec{D}_{ij}\cdot \vec {S}_i \times \vec {S}_j 
 $$
 While the Heisenberg exchange coupling  is symmetric under exchange of spins (for real couplings) and gives rise to collinear (parallel or antiparallel) magnetic ordering, the DMI is antisymmetric under exchange of spins and induces canting of spins which can lead to noncollinear magnetic order.  
 The competition between the canting of spins due to DMI and the parallel/antiparallel magnetic ordering due to the Heisenberg exchange coupling provides a mechanism for
 frustration in the magnetic system, independent of geometry of the lattice.
A lattice-form-independent microscopic theory of the MEE, the so called spin current or inverse DM  mechanism, has been successfully used to explain the origin of  magnetically induced  ferroelectricity in many spiral magnets~\cite{knb}.
 The electric  polarization operator $\vec{P}_{ij}$ has been shown to take the form~\cite{knb}:
 $$\vec{P}_{ij} \sim \gamma \hspace{1mm}\hat{e}_{ij} \times (\vec{S}_i\times\vec{S}_j),$$ 
 where $\hat{e}_{ij}$ is the unit vector along the bond joining spins at sites $i$ and $j$,
and $\vec{S}_i\times\vec{S}_j$ is the spin current.

It is well known that external magnetic fields drive quantum phase transitions in the Heisenberg spin model. In previous studies, we showed that the spin-orbit coupling or the DMI can also drive phase transitions~\cite{pradeepPRB1,pradeepPRB2}. Interpreting the DMI as an electric field coupling to the polarization operator, we observed two types of MEE.
There is a `trivial' MEE with noncollinear spin ordering and a consequent finite electric polarization induced by the external electric field or DMI, and tunable by magnetic fields. The other type of MEE we observed is a `nontrivial' MEE, where a transverse magnetic field induced a noncollinear ordering of spins and a consequent staggered electric polarization.
What is common in both these cases is that the magnetic/electric order was produced by an external electric/magnetic field, and vanished in the absence of the field. A natural question to ask is - whether spontaneous magnetic and/or electric order can exist in such systems in absence of external fields (and without explicit breaking of parity through lattice geometry).
We have partially answered this question in our earlier studies~\cite{pradeep2016,pdn2016} where we considered 
a spin-$1/2$ XX chain with additional three-spin interactions: one of the type $S_n^{x}S_{n+1}^{z}S_{n+2}^{x}+S_n^{y}S_{n+1}^{z}S_{n+2}^{y}$ denoted $XZX+YZY$, along with one of the type $S_n^{x}S_{n+1}^{z}S_{n+2}^{y}-S_n^{y}S_{n+1}^{z}S_{n+2}^{x}$, denoted $XZY-YZX$, and showed that there is a phase with spontaneous magnetic and electric order, i.e. in the absence of any external electric or magnetic fields.

Multi-spin interactions were first studied in the context of integrable low-dimensional spin models~\cite{suzukiPLA1971, suzukiPTP1971, baxterwu}. The purpose of study has also been the exploration of the unique phases that these models host and the transitions between them~\cite{penson1982}. Three-spin interactions, specifically, are the simplest nontrivial case and have been investigated in a variety of contexts: critical behaviour of these models and the relation to the underlying conformal field theory~\cite{penson1988}, exact solution in presence of Dzyaloshinskii-Moriya interaction~\cite{gottlieb}, quantum phase transitions~\cite{titvinidze2003, lou2004}, dynamical properties~\cite{krokhmalskii2008}, magnetocaloric effect~\cite{topilko2012} and  magnetoelectric effect~\cite{menchyshyn2015}.

The model with three-spin interactions of the $XZX+YZY$ and $XZY-YZX$ types ~\cite{suzukiPLA1971, suzukiPTP1971} has been shown to exhibit a rich phase diagram~\cite{titvinidze2003,lou2004}. In the presence of three-spin interactions, there is a Lifshitz transition, with a change in the Fermi sea topology which in this one dimensional system is characterized by change in the number of Fermi points from two to four at a certain critical three-spin coupling strength. The model with three-spin interactions of the $XZX+YZY$ type was shown to have %a phase with
 spontaneous magnetization~\cite{titvinidze2003} in one of its phases, while the model with three-spin interactions of the $XZY-YZX$ type was shown to have spontaneous scalar chiral order~\cite{lou2004}, in a certain parameter regime.  More recently, a nontrivial MEE was reported in these models~\cite{menchyshyn2015}, where an external electric field influenced the magnetic order in absence of an external magnetic field, and a magnetic field influenced the ferroelectric order in the absence of an electric field. In the absence of external fields however, they showed that the total electric polarization is zero.

In this paper, we revisit the model and show that \textit{when both the types} $XZX+YZY$ and $XZY-YZX$ are present, the model exhibits 
exotic ground state behaviour, with a novel higher dimensional magnetoelectric order
even in the absence of external fields. 
In earlier studies, we showed that when both the types $XZX+YZY$ and $XZY-YZX$ are present, the model shows a phase with spontaneous magnetic and electric order~\cite{pradeep2016}. 
In this paper, we elaborate on the rich properties of this phase. We show that even though the model is one-dimensional, a higher dimensional magnetoelectric order emerges due to a nontrivial topological connectivity.
Specifically, we find a novel orbital antiferroelectric order~\cite{pdn2016}. We show that in this novel phase, in addition to a spontaneous magnetization and scalar chirality,  there are non-vanishing plaquette scalar and vector chiral orders, which arise due to non-coplanar ordering of spins in a plaquette of three adjacent spins. We characterize the higher dimensional magnetoelectric order by scalar and vector toroidal order parameters. Further, we probe the bipartite entanglement entropy in the presence of both types of three-spin interactions. We observe an interesting non-trivial dependence of the entanglement entropy on the strength of the $XZY-YZX$ interaction, in the presence of the $XZX+YZY$ interaction.

The plan of the paper is as follows. We begin in Sec.~\ref{sec:model} by defining the model, and discuss the various symmetries of the model. We also map the model to the corresponding free spinless fermion model via the Jordan-Wigner transformation and define the various quantities of interest. This is followed by our results in Sec.~\ref{sec:res}. where we begin by first discussing the ground state phase diagram of the model. We then describe our results for the various physical quantities of interest like the magnetization, polarization, scalar and vector chiral orders, plaquette vector chirality, etc. We show the existence of a novel phase with an `orbital' antiferroelectricity which arises due to the non-trivial topological connectivity. In Sec.~\ref{sec:EE}, we describe the non-trivial topological order by studying the behaviour of the bipartite entanglement entropy. Sec.~\ref{sec:conc} summarizes our results.

\section{\label{sec:model}The Model}
   The Hamiltonian for a N-site spin-$1/2$ $XX$-chain with three-spin interactions of the $XZX+YZY$ and $XZY-YZX$ types can be written as \cite{suzukiPTP1971}:
\begin{gather}
H_0 = H_{XX} + H_A + H_B, \text{where}\nonumber \\
H_{XX} = {\sum_{n=1}^N}\,J(S_n^x S_{n+1}^x + S_n^y S_{n+1}^y), \nonumber \\
H_A = {\sum_{n=1}^N}\, A (S_n^{x}S_{n+1}^{z}S_{n+2}^{x}+S_n^{y}S_{n+1}^{z}S_{n+2}^{y}), \text{and} \nonumber \\
H_B = {\sum_{n=1}^N}\,B (S_n^{x}S_{n+1}^{z}S_{n+2}^{y}-S_n^{y}S_{n+1}^{z}S_{n+2}^{x})
\label{eq:ham0}
\end{gather}
Here, $A$ and $B$ denote the strengths of the three-spin exchange interactions of the $XZX+YZY$ and $XZY-YZX$ types respectively. We also assume periodic boundary conditions $\vec S_{N+1} = \vec S_{1}$. 
A uniform external magnetic field $h_z$ in the $z$ direction couples as usual to the total spin operator
$ S^z= {\sum_{n=1}^N} S_n^z$ giving rise to the additional Zeeman  interaction in the Hamiltonian: $-h_z S^z$. We also consider an external electric field which couples, within the KNB mechanism\cite{knb}, to the electric polarization operator, $\hat{\vec{P}}_{n,m} = {\vec e}_{nm}\times \hat{\vec{j}}_{n,m}$, where $\hat{\vec{j}}_{n,m}$ is the spin current operator between sites $n$ and $m$, given by 
\begin{equation}
{\vec {\hat j}}_{n,m} = {\vec S}_n \times {\vec S}_{m}
\label{eq:sc1}
\end{equation}
and all material-dependent constants have been absorbed in the corresponding electric field. The spin current operator can be obtained from the equation of motion for the $S_n^z$ operator as \cite{pradeep2016, menchyshyn2015}:
\begin{gather}
\frac{dS_n^z}{dt} = -i \left[ S_n^z, H_{XX} +  H_A + H_B \right] \nonumber \\
= -i [ S_n^z, H_{XX} ] - i \left [S_n^z, H_A + H_B \right]
\label{eq:sc2}
\end{gather}
The first commutator is as given by: 
\begin{gather}
-i [S_n^z,H_{XX}] = -\left(\hat{j}_{n+\frac{1}{2}}^z - \hat{j}_{n-\frac{1}{2}}^z\right) = -\text{div}\, \hat{j}_n^z
\label{eq:j1}
\end{gather}
where $\hbar$ has been set to $1$, and
\begin{equation}
 \hat{j}^z_{n+\frac{1}{2}} = \hat{j}^z_{n,n+1} = J (S_n^x S_{n+1}^y - S_n^y S_{n+1}^x)
\end{equation}
while the second commutator can be obtained as:
\be
-i \left[S_n^z, H_A + H_B \right] = -\frac{\hat{\j}_{n+1}^{z} - \hat{\j}_{n-1}^{z}}{2} = -\text{div}\,\, \hat{\j}_n^{z}
\label{eq:j2}
\ee
where
\begin{gather}
\hat{\j}_{n-1}^{z} = 2A (S_{n-2}^{x}S_{n-1}^{z}S_{n}^{y}-S_{n-2}^{y}S_{n-1}^{z}S_{n}^{x}) \nonumber \\ - 2B (S_{n-2}^{x}S_{n-1}^{z}S_{n}^{x}+S_{n-2}^{y}S_{n-1}^{z}S_{n}^{y})
\label{eq:j2def}
\end{gather}
\textit{Note the difference} between the symbol $\hat{j}$ representing the spin current originating from the two-spin interaction and the symbol $\hat{\j}$ representing the spin current originating from the three-spin interactions, \textit{e.g.} in Eqs.~\ref{eq:j1} to~\ref{eq:j2def}. The electric polarization operator ${\hat P}^y$ can therefore be written as:
\begin{equation} 
{\hat P}^y = {\hat P}_1^y + {\hat P}_2^y;\,\, {\hat P}_1^y = -\sum_n\hat{j}_{n+1}^z; \,\, {\hat P}_2^y = - \sum_n \hat{\j}_{n+1}^{z}
\label{eq:p1p2}
\end{equation}

Hence, the Hamiltonian in the presence of an external electric field $E_y$ $(=E)$ along the $y$ direction and a magnetic field $h_z$ along the $z$ direction is given as:
\begin{gather}
\mathcal{H} = H_0 - E {\hat P}^y  - h_z S^z \nonumber \\
= {\sum_{n=1}^N}\, [ J (S_n^x S_{n+1}^x + S_n^y S_{n+1}^y)  \nonumber \\ + E (S_n^x S_{n+1}^y -  S_n^y S_{n+1}^x ) -  h_z S_n^z \nonumber \\
+ (A -2B E) (S_n^{x}S_{n+1}^{z}S_{n+2}^{x}+S_n^{y}S_{n+1}^{z}S_{n+2}^{y}) \nonumber \\
+  (B + 2 A E) (S_n^{x}S_{n+1}^{z}S_{n+2}^{y}-S_n^{y}S_{n+1}^{z}S_{n+2}^{x})\,] 
\label{eq:ham_3spin}
\end{gather}

Alternately, redefining $A -2B E$ as $A$ and $B + 2 A E$ as $B$, we can express the Hamiltonian as:
\begin{gather}
\mathcal{H} = \sum_{n=1}^{N-1} ( \frac{1}{2}{\tilde J}e^{i \phi^1} S_n^+ S_{n+1}^- + h.c ) -  h_z \sum_{n=1}^{N} S_n^z \nonumber \\
+ \sum_{n=1}^{N-1} (\frac{1}{2}{\tilde A} e^{ i \phi^2} S_{n-1}^+ S_{n+1}^- + h.c )S_n^z %\nonumber \\ %+ at the end removed
%+ \sum_{n=1}^{N-1} ( {\tilde \gamma} e^{i \phi^3} S_n^+ S_{n+1}^+ + h.c ) 
%\sum_{n=1}^{N-1} ( {\tilde \delta} e^{i \phi^4} S_{n-1}^+ S_{n+1}^+ + h.c ) S_n^z \nonumber \\
%+ ({\tilde J_N} e^{i \phi^1} S_N^+ S_1^- + h.c )
%+ ( {\tilde A} _n e^{i \phi^2} S_{N-1}^+ S_{1}^- + h.c) S_N^z
\label{eq:Ham2}
\end{gather}
where  $\tilde J = \sqrt{J^2 +E^2},\,\,\, \phi^1 = \tan^{ -1} \frac{E}{J}, \,\, \tilde A = \sqrt{A^2 +B^2}, \,\, \phi^2 = \tan ^{ -1} \frac{B}{A}$, and the operators $S_n^{\pm}$ are defined as $S_n^{\pm} = S_n^x \pm i S_n^y$, hence the absence of boundary terms in the above equation.

It is useful to analyze the symmetries of the Hamiltonian before studying the solutions\cite{krokhmalskii2008}. Under a local rotation of the spin operators by angle $\phi_n$  about the $z$ axis, the spin operators transform as: 
\begin{gather}
S_n^+ \rightarrow {S'}_n^+ =  e^{i \phi_n} S_n^+ \nonumber \\  S_n^- \rightarrow {S'}_n^+ =  e^{-i \phi_n} S_n^- \nonumber \\
S_n^z \rightarrow {S'}_n^z = S_n^z
\label{eq:gauge}
\end{gather}
and defining 
\begin{equation}
\psi_n \equiv \phi_{n+1}- \phi_n ; \quad { \theta}_n \equiv \phi_{n+1}- \phi_{n-1},  
\end{equation}
the Hamiltonian (Eq.~\ref{eq:Ham2}) transforms as: 
\begin{gather}
\mathcal{H} = \sum_{n=1}^{N-1} ( \frac{1}{2}{\tilde J}e^{i (\phi^1 + \psi_n) } S_n^{'+} S_{n+1}^{'-} + h.c ) -  h_z \sum_{n=1}^{N} S_n^z \nonumber \\
+ \sum_{n=1}^{N-1} (\frac{1}{2}{\tilde A} e^{ i (\phi^2 + \theta_n)} S_{n-1}^+ S_{n+1}^- + h.c )S_n^z
\label{eq:rot_Ham2}
\end{gather}

Thus, we can transform the $J, E,A,B$ model into one with modified parameters $J', E',A',B'$ by the gauge rotation (Eq.~\ref{eq:gauge}).
The $h_z$ term does not change under the rotation.
We consider now the interdependence of $E$, $A$, and $B$ through the gauge symmetry defined above.
It is useful to define some specific cases of models of interest as:\\
 (i) $A=0, B =0$ ($JE$ model), \\
 (ii) $E=0, B=0$ ($JA$ model), \\
 (iii) $E=0, A=0$ ($JB$ model), \\
 and (iv) $E=0, A,B \neq 0$ ($JAB$ model),which corresponds to a model with no external fields: $h_z, E = 0$.
 
 Starting from the $JE$ model, we can eliminate the DMI term or equivalently the phase $\phi^1$ in Eq.~\ref{eq:Ham2} by a  rotation of the spin operators about the $z$ axis by an angle $ \phi_n = -n \phi^1=  -n \tan^{-1} \frac{E}{J}$ leading to a  Hamiltonian  with a modified Heisenberg exchange coupling $J'= \sqrt{J^2 +E^2}$. Starting from the $JA$ model, it is possible to eliminate the three-spin $A$ interaction term through a gauge rotation of the spin operators about the $z$ axis by $\phi_n = n \pi/4$ and obtain a model with $J',B',E'$ terms, where $J' = \frac{1}{\sqrt 2} J,\, E'= -\frac{1}{\sqrt 2} J,\, B' = A$. Similarly, the $J, B$ model can be rotated by a gauge rotation $\phi_n = - n \pi/4$ into a $J',A',E'$ model, where $J' = \frac{1}{\sqrt 2} J, E'= -\frac{1}{\sqrt 2} J, A' = \frac{1}{\sqrt2} B$.  
 Consider now the  $JAB$ case where both $A$ and $B$ are present but no $E$. Then $\phi^1=0$ but $\phi^2 = \tan^{-1}\frac{B}{A}$
In this case we can eliminate either $A$ or $B$ by a gauge rotation but at the cost of introducing an $E'$. For example, we can eliminate the phase $\phi^2$ or equivalently the $B$ term through  a rotation of the spin operators by $\phi _n = -\frac{1}{2} n\phi ^2$.  But this rotation modifies the phase $\phi^1$ to  $\phi'^1 = - \frac{1}{2} \phi^2$.  Thus the Hamiltonian gets rotated into one with modified parameters: $ J ' = J \cos \left( \frac{1}{2}  \tan^{-1} \frac{B}{A}\right)$, $E '=  - J \sin \left(\frac{1}{2}  \tan^{-1} \frac{B}{A}\right), A' = \sqrt {A^2 +B^2}, B'=0$.  Thus we can rotate into a model with  only $ J', E', A'$ terms. Similarly, if we rotate the spin operators by $\phi _n = -\frac{1}{2} n\phi ^2 +  n \pi/4 $, then we can eliminate the $A$ term and obtain a  Hamiltonian with modified parameters:
$ J ' = J \cos \left(\pi/4 - \frac{1}{2}  \tan^{-1} \frac{B}{A}\right)$, $ E '=  J \sin \left(\pi/4 - \frac{1}{2}  \tan^{-1} \frac{B}{A}\right)$, $A'=0$, $B' = \sqrt {A^2 +B^2}$. Thus we get a model with only $J', E', B'$ terms.  Again, in both these cases, the $E'$ terms cannot be further gauged away to get pure $J,A$ or $J,B$ models.
We note here that $E'$ is not arbitrary but fixed by the gauge transformation and cannot be further gauged away.  We emphasize here that the $J,A,B$ model cannot be transformed into pure $J',A'$ or $J',B'$ models without any additional DMI term.
(For the sake of completeness, when we consider the $JEAB$ case, where there is an additional $E$ present, we can eliminate $E$ by making the rotation: $\phi_n= -n \phi ^1$ which transforms the model to one with modified $J,A,B$ terms. Or we can make a rotation transformation like in cases (ii) and (iii) and eliminate either $A$ or $B$, in which case we end up with  modified $J, B, E$ or $J,A,E$ models.) 

The Hamiltonian (Eq.~\ref{eq:ham_3spin}) can be converted into a fermionic tight binding Hamiltonian by a Jordan-Wigner transformation of the spin operators:
\begin{gather}
c_n \equiv \text{exp}\left[ \pi i \sum_{m=1}^{n-1} S_m^+ S_m^- \right] S_n^-;\nonumber \\
c_n^{\dagger} \equiv S_n^+ \text{exp}\left[ -\pi i \sum_{m=1}^{n-1} S_m^+ S_m^- \right]
\label{eq:JW_c1}
\end{gather}
It follows that $\displaystyle{c_m^{\dagger}c_m = S_m^+ S_m^-}$ 
leading to the inverse transformations:
\begin{gather}
S_n^- = \text{exp}\left[ -\pi i \sum_{m=1}^{n-1} c_m^{\dagger} c_m \right] c_n; \nonumber \\
S_n^+ = c_n^{\dagger} \text{exp}\left[ \pi i \sum_{m=1}^{n-1} c_m^{\dagger} c_m \right]
%\quad S_i^z = c_i^{\dagger} c_i - 1/2
\label{eq:JW_S1}
\end{gather}
and the spin operator $S_n^z$ is expressed as:
\begin{equation}
S_n^z = S_n^+ S_n^ - -\frac{1}{2} = c_n^{\dagger} c_n - \frac{1}{2}
\label{eq:JW_Sz}
\end{equation}
Using these we convert the Hamiltonian (Eq.~\ref{eq:ham_3spin}) into a tight binding Hamiltonian for spinless fermions with nearest neighbour (NN) and next nearest neighbour (NNN) hopping:
\begin{gather}
H_0 = \sum_{n=1}^N [\frac{J}{2}(c_n^{\dagger}c_{n+1} + c_{n+1}^{\dagger}c_{n}) 
- h_z (c_n^{\dagger}c_{n}-\frac{1}{2}) \nonumber \\ -  
\frac{E}{2i} (c_n^{\dagger}c_{n+1} - c_{n+1}^{\dagger}c_{n}) \nonumber \\
-\frac{(A - 2 B E)}{4}(c_n^{\dagger}c_{n+2} + c_{n+2}^{\dagger}c_{n}) \nonumber \\ - 
\frac{i (B + 2 A E)}{4} (c_n^{\dagger}c_{n+2} - c_{n+2}^{\dagger}c_{n})]
\end{gather}
A subsequent  Fourier transformation of  the fermion operators to momentum space  ($a$ is the lattice constant) results in a diagonalized Hamiltonian for non-interacting spinless fermions:
\begin{equation}
\mathcal{H} = \sum_k (\varepsilon_k)c_k^{\dagger}c_k +\frac{Nh_z}{2}
\label{eq:fermion_ke}
\end{equation}
with the sum over the wave vectors $k$ being restricted to the first Brillouin zone: $\pi/a \leq k <\pi/a$. 
The single particle energies $\varepsilon_k$ satisfy the dispersion relation:
\begin{gather}
\varepsilon_k  = -h_z + \epsilon_k \nonumber \\
= -h_z +J\,\text{cos}(ka)- E\,\text{sin}(ka) \nonumber \\ -\frac{ A - 2 E B}{2}\,\text{cos}(2ka)+\frac{B + 2 E A}{2}\,\text{sin}(2ka)
\label{eq:disp}
\end{gather}
It can be seen from the above dispersion relation that the three-spin exchange interactions and the external electric field or the DMI, $E$, modify the single particle excitation energy  while the external longitudinal magnetic field, $h_z$, acts like a chemical potential.
In the thermodynamic limit ($N\rightarrow\infty$), the ground state configuration of the system has all $k$-states with $\varepsilon_k\leq0$ occupied, and those with $\varepsilon_k>0$ unoccupied. For a one-dimensional system, the Fermi surface is zero-dimensional; the Fermi energy occurs at the points $\varepsilon_k=0$ and the wave-vectors, $k_F$, at which the energy $\varepsilon_k$ vanishes, are the Fermi points. 

The partition function for the non-interacting fermion system is readily obtained as:
\begin{equation}
Z_{N}(T, A, B,h_z, E) = \text{Tr} e^{-\beta \mathcal{H}} = e^{\frac{Nh_z}{2k_{B}T}}\prod_k (1 + e^{\frac{-\varepsilon_{k}}{k_{B}T}})
\end{equation}
Henceforth, we set $k_B=1$ and the lattice constant $a=1$.
The Helmholtz free energy per site  is given as:
\begin{gather}
f_{N}(T, A, B,h_z, E) \equiv  -\lim_{N\to \infty} \frac{T\hspace*{1mm}\text{ln}Z_{N}(T, A, B,h_z, E)}{N}\nonumber \\= const. + \frac{1}{2\pi}\int_{-\pi}^{\pi} dk \left( \varepsilon_{k} + T\hspace*{1mm}\text{ln} n_{k}\right)
\label{eq:f}
\end{gather}
where $n_k \equiv 1/(e^{\varepsilon_{k}/T}+1)$ is the fermion occupation number which at $T = 0$, has the expectation value $1$ for the occupied states and $0$ for the unoccupied states.
One can now calculate the various thermodynamic quantities from the derivatives of the free energy, \textit{e.g.} the thermodynamic entropy $s_{}(T, A, B,h_z, E) $ per spin is obtained as 
\begin{gather}
s_{ }(T, A, B,h_z, E) \equiv - \frac{\partial f(T, A, B,h_z, E)}{\partial T}\nonumber \\= - \frac{1}{2\pi}\int_{-\pi}^{\pi} dk \left(\text{ln}\,n_{k} + \frac{\varepsilon_{k}}{T}\,e^{\varepsilon_{k}/T}n_{k}\right)
\end{gather}
In the rest of the chapter, we restrict to the case where there are no external fields, i.e. $h_z=0$, $E=0$. We also discuss only the ground state properties.

\section{Ground state properties}\label{sec:res}
\subsection{\label{sec:GSPD}Ground state phase diagram}
In this section, we describe the various ground state phases and their dependence on the three-spin interactions $A$ and $B$. The phase diagrams for the case where only one of the three-spin interactions is present have been discussed previously~\cite{titvinidze2003, lou2004}.
We show here that presence of both $A$ and $B$ types of three spin interactions leads to qualitatively new physics.
\begin{figure}[!htpb]
\includegraphics[width=3.0in]{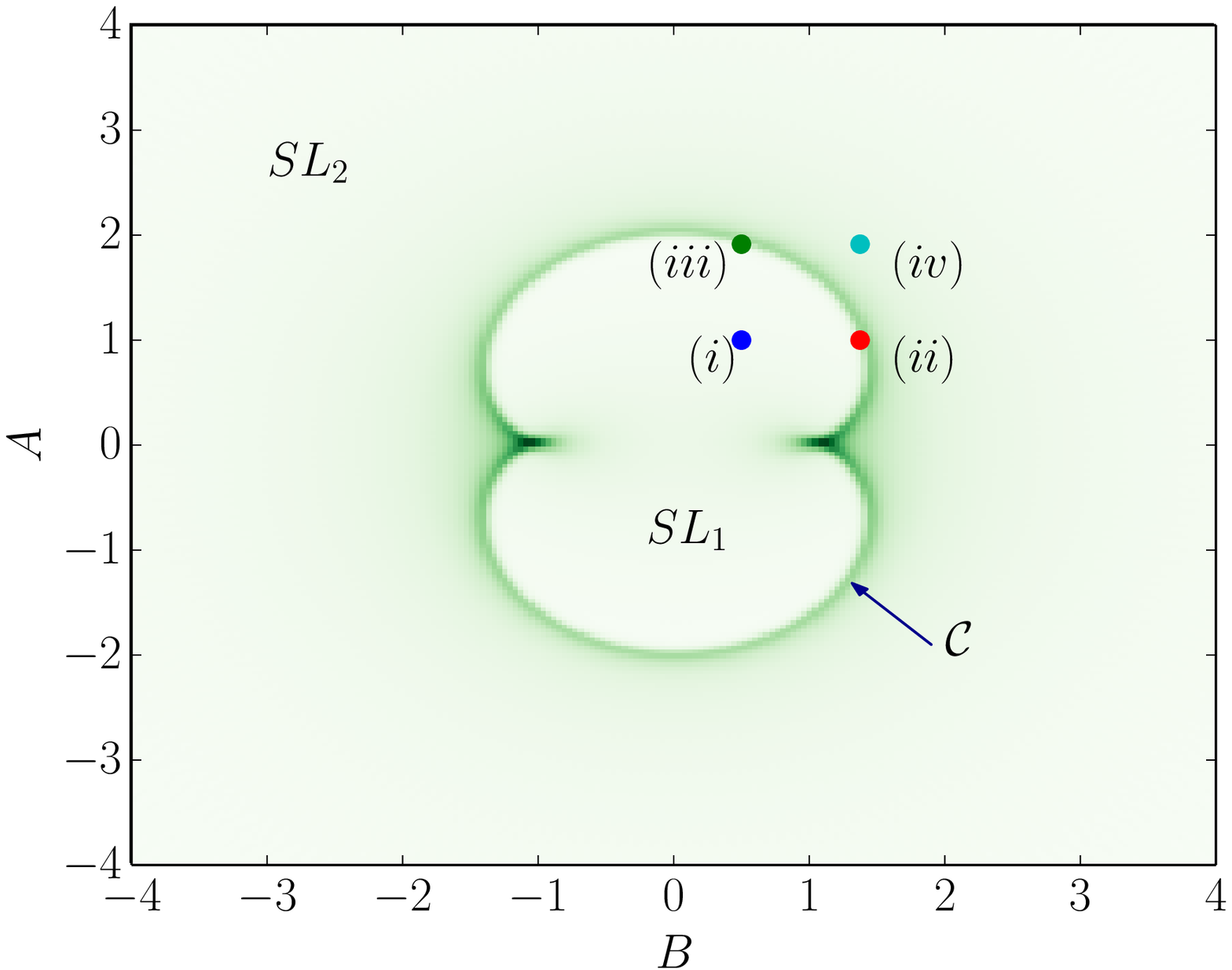}{(a)}
\includegraphics[width=3.0in]{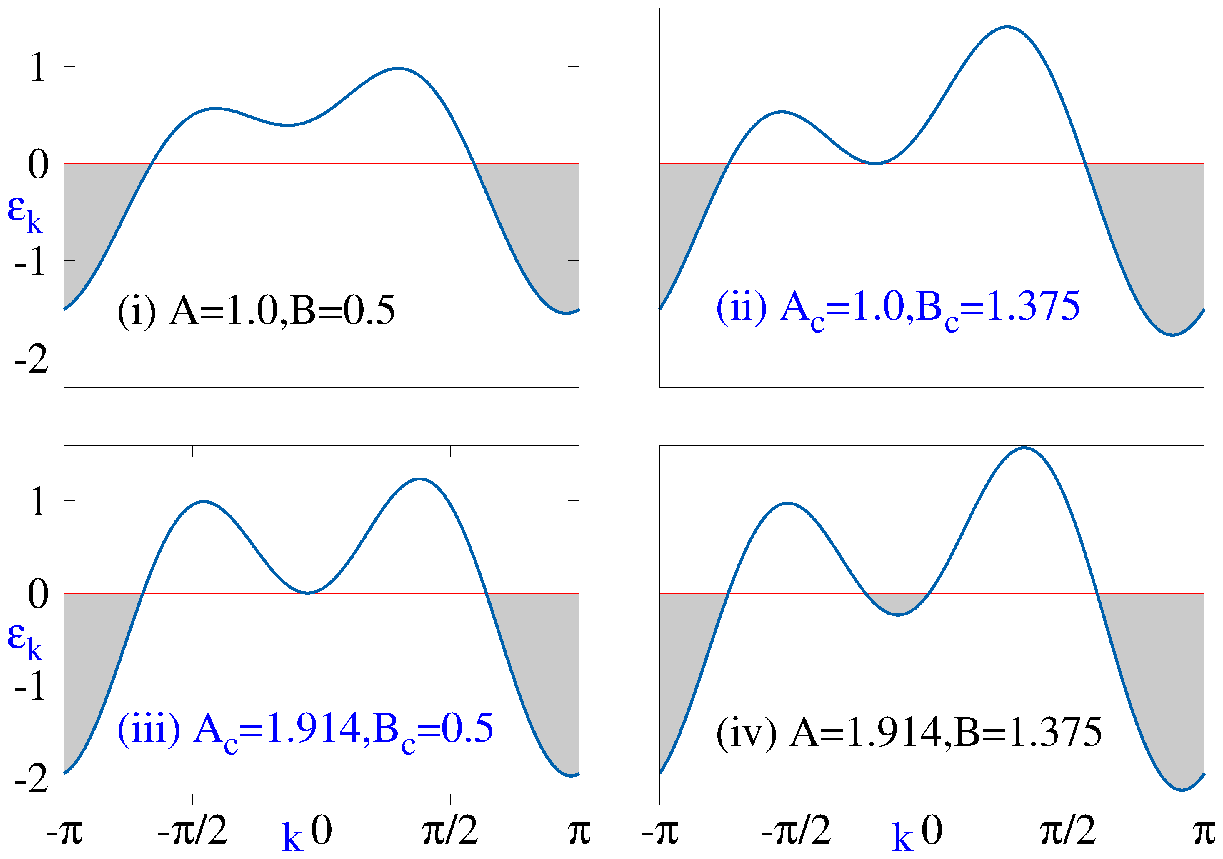}{(b)} 
\caption[(a) Schematic $B-A$ phase diagram and (b) dispersion plots in the presence of both $A$ and $B$.]{(Upper panel) A schematic phase diagram marking the transition (critical curve $\mathcal{C}$) between the two phases $SL_1$ and $SL_2$. (Lower panel) The energy-dispersion $\varepsilon$ versus wave vector $k$ in the presence of both three-spin interactions. The subpanels (i),(ii),(iii) and (iv) of the lower panel are dispersion plots at the corresponding points marked in upper panel.}
\label{fig:disp2}
\end{figure} 
We show, in the upper panel of Fig.~\ref{fig:disp2}, a schematic phase diagram for the general case when both $A$ and $B$ are present.
The critical transition curve, denoted by $\mathcal{C}$, identifies a Lifshitz transition~\cite{lifshitz1960}, characterized by a change in the topology of the Fermi sea, and separating two different phases Spin-liquid-I ($SL_1$) and Spin-liquid-II ($SL_2$) in the $B-A$ plane~\cite{titvinidze2003}. The $SL_1$ phase is characterized by a simply connected $1d$ Fermi sea, while the $SL_2$ phase is characterized by a $1d$ Fermi sea made of two disjoint pieces. The boundaries of the Fermi sea are called Fermi points.
For the parameter regime $A <A_c; B<B_c $, lying within the critical curve, there are two Fermi points as shown through the plots for the energy-momentum dispersion (Eq.~\ref{eq:disp} in subpanel (i)) of the lower panel of Fig.~\ref{fig:disp2} for representative values, while for the parameter region lying outside the critical curve, there are four Fermi points as can be seen from subpanel (iv) of the lower panel.
(In the case of a single three-spin interaction, the values of the Fermi points can be obtained analytically\cite{titvinidze2003,lou2004}.)

Subpanels (ii) and (iii) show the dispersion at points on the critical curve. It can also be seen that when both $A$ and $B$ are non-zero, the dispersion is not invariant under parity. The Fermi sea is a connected region for $A <A_c,\, B<B_c$, while for $A >A_c, \, B>B_c$, the Fermi sea becomes disconnected into two separate pieces. A second order QPT occurs at each critical point lying on the critical curve with the ground state stiffness showing a divergence.
The regions interior to the critical curve (with two Fermi points) and exterior to the critical curve (with four Fermi points) are both gapless and believed to be described by conformal field theories with central charge $c=1$ and $c=2$ respectively~\cite{titvinidze2003}.
The spin correlation functions have power law behaviour in both phases with differing exponents which depend now on both $A$ and $B$. 

\begin{figure}[htpb!]
\includegraphics[width=3.0in]{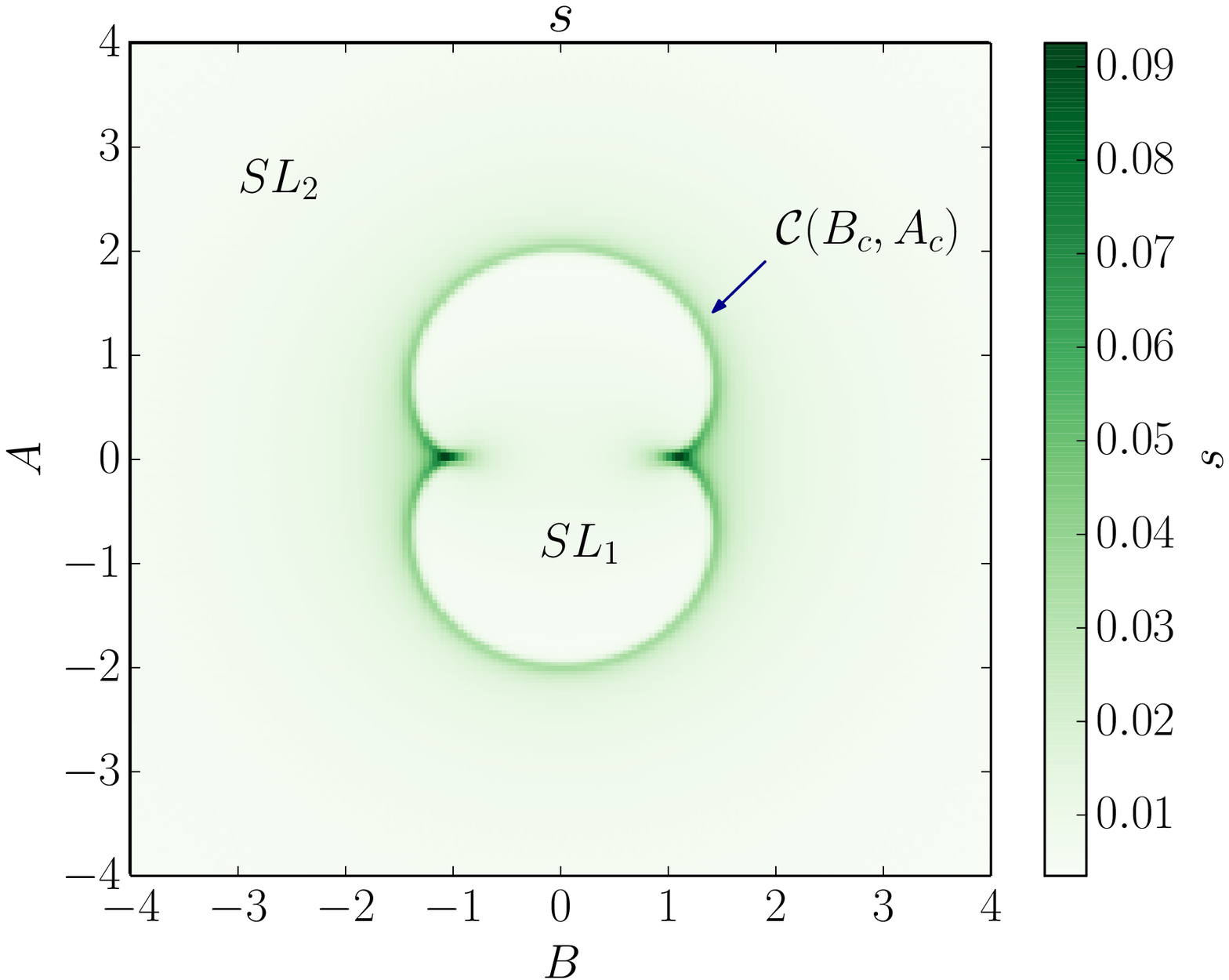}{(a)}
\includegraphics[width=3.0in]{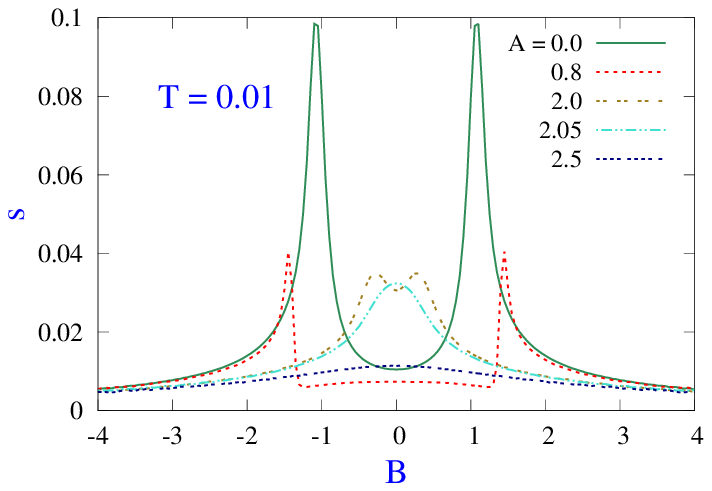}{(b)}
\caption[The $B-A$ dependence of the entropy.]{(a) The $B-A$ dependence of the entropy, $s$. (b) The $B$ dependence of the entropy for different fixed values of $A$. Comparing the height of the entropy peaks, the peak height for $B_c = \pm 1, A_c=0$ is approximately $0.1$, whereas at the QCP, say, at $B_c=0, A_c=2.0$, the peak height is $\sim0.035$.}
\label{fig:S_phases}
\end{figure}

One can in fact obtain the ground state phase diagram from the zero temperature thermodynamic entropy.
In the upper panel of Fig.~\ref{fig:S_phases}, we show the $B-A$ dependence of the zero temperature thermodynamic entropy, $s$, per unit spin and in panel (b) of the figure, we plot the $B$ dependence of the entropy for fixed $A$ values, i.e. along horizontal segments of the panel. We can see that the entropy is maximum along the critical curve $\mathcal{C}(B_c, A_c)$. $\mathcal{C}(B_c, A_c)$ is a nephroid, given by the equation $(B_c^2 + A_c^2 - 1)^3 = 27A_c^2/4$.
Further, we observe that the critical curve or equivalently the maximal entropy curve is not smooth at the points where the global maxima occur; the curvature diverges at these points. Also, we find the ratio of the entropy at the non-smooth points ($A_c=0$, $B_c=\pm 1$) to that at other QCPs along the critical curve $\mathcal{C}$ to be approximately $3:1$ as can be seen from the same figure panel.

In the next subsection we describe various physical properties of the model, focussing on the magnetization, electric polarization, scalar chirality and vector chirality properties.
\subsection[Local orders]{\label{sec:order}Magnetization, electric polarization, scalar chirality and vector chirality}
\begin{figure}[ht!]
\centering
\includegraphics[width=3.0in]{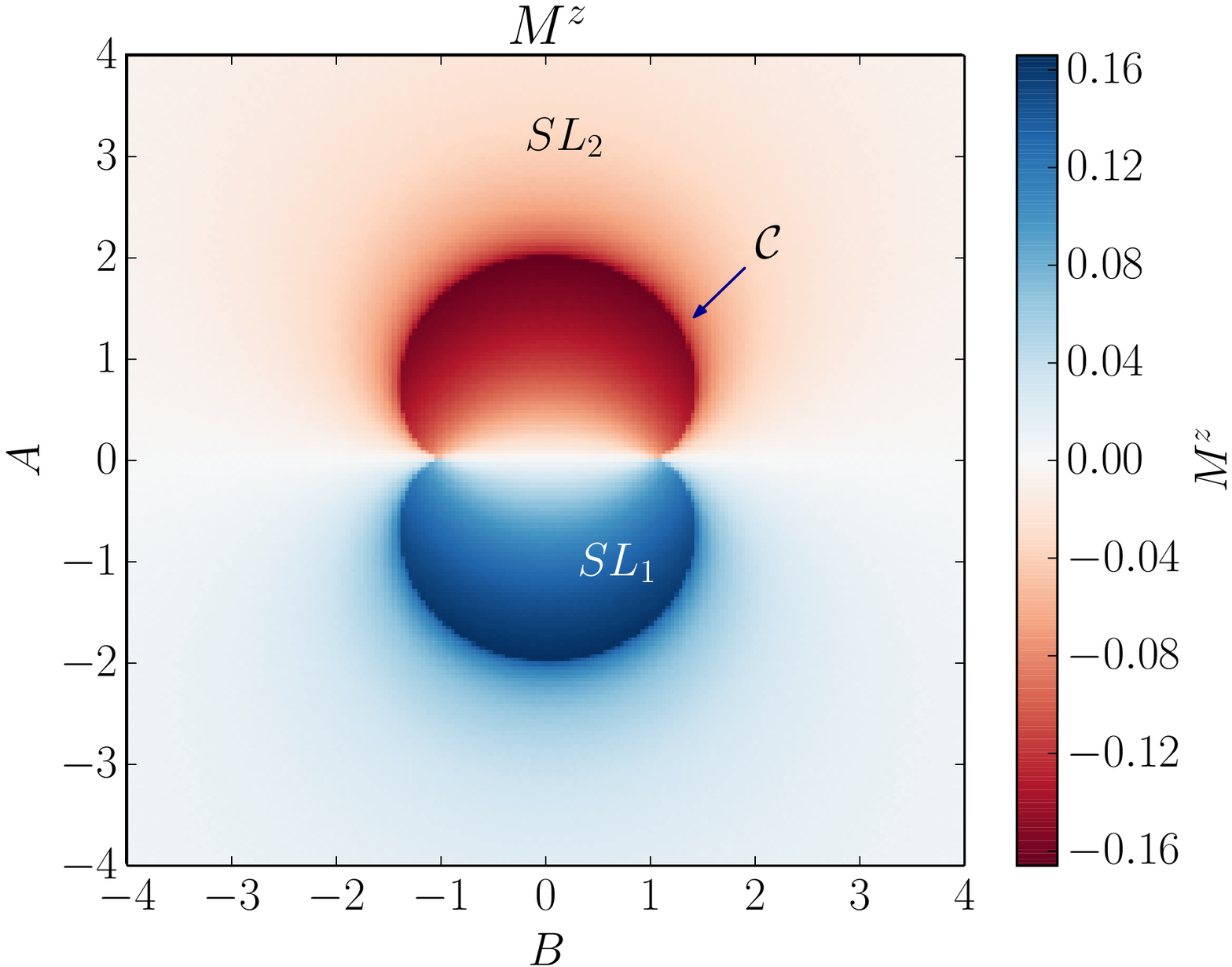}{(a)}
\includegraphics[width=3.0in]{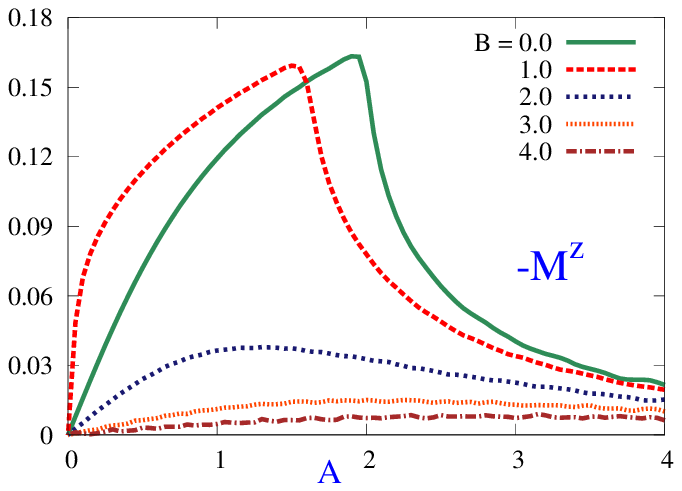}{(b)}
\includegraphics[width=3.0in]{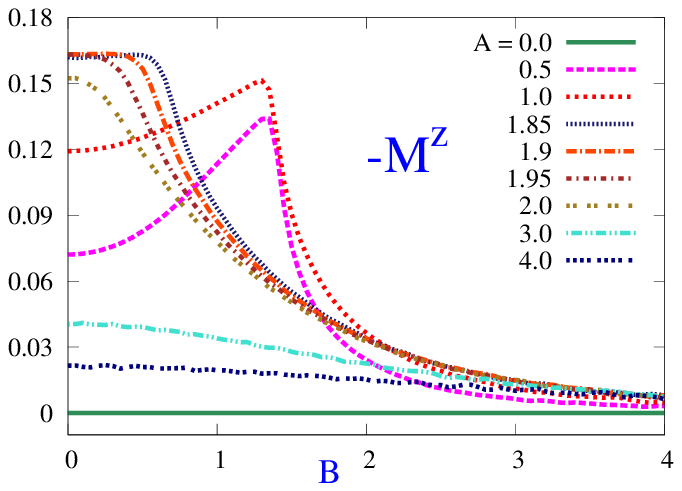}{(c)}
\caption[The $B-A$-dependence of the magnetization per spin $M^z$.]{\label{fig:mag} (a)The $B-A$-dependence of the magnetization per spin $M^z$ at $T=0.001$. 
(b) The $A$-dependence of $M^z$ for different values of $B$, and (c) the $B$-dependence for different values of $A$, with $T=0.01$ for both these panels. We see that inside the $SL_I$ phase and in the vicinity of its boundary, $M^z$ is zero in the absence of $A$, but nonzero otherwise.} 
\end{figure}

While collinear magnetically ordered phases are typically characterized by the uniform or staggered magnetization, several different order parameters like the scalar chirality, vector chirality, plaquette vector chirality etc. have been introduced to describe noncollinear magnetically ordered systems.
We discuss in this subsection the behaviour of the magnetization, electric polarization, scalar chirality, vector chirality, plaquette  vector chirality etc. 
The (uniform) magnetization per unit spin, $M^z$,  can be obtained from the free energy (Eq.~\ref{eq:f}) as:
\begin{gather}
M^{ z} = \lim_{N\to \infty} \frac{1}{N}\sum_{n=1}^{N} \langle S_{n}^{z}\rangle \nonumber \\
= - \frac{\partial f_{N}}{\partial h_z} = \frac{1}{2\pi}\int_{-\pi}^{\pi} dk \left(n_{k} - \frac{1}{2}\right)
\end{gather}
where we have suppressed the arguments of $f_N$. In Fig.~\ref{fig:mag}(a), we show the $A-B$ dependence of the magnetization. The behaviour of the magnetization as a function of $A(B)$ for fixed values of $B (A)$ is shown in panels (b,c).
\noindent It can be seen from Fig.~\ref{fig:mag} that the magnetization vanishes if $A=0$. The maximum value of the magnetization occurs along the critical curve $\mathcal{C}$.
The magnetization decays for very large $A$ and $B$ values.
For non-vanishing $A$, there is a finite magnetization with a non-monotonic $A$ dependence.
In the presence of a finite $B$, the magnitude of the magnetization decreases from that in the absence of $B$ as can be seen from Fig.~\ref{fig:mag}(a,b). From the $B$ dependence of the magnetization shown in Fig.~\ref{fig:mag}(c), we can again see that the magnetization becomes finite as soon as $A$ is finite and then reaches a maximum at the critical value. We also note that the magnetization is odd with respect to $A$ and even with respect to $B$.  

   We next study the chiral properties of the model. 
The scalar chirality operator $\hat \chi_{lmn}$ is defined as:
\be
\hat \chi_{l m n} = \vec S_l . \vec S_{m} \times\vec S_{n}
\ee
 We also introduce the plaquette scalar chirality $\chi$ defined as
 \be
 \chi =  \lim_{N\to \infty} \frac{1}{N} \sum_n  \langle \hat \chi_{n,n+1, n+2}\rangle
 \ee
 and is given as:
 \be
 \chi = -O_B (1 - 4 O_J) + 2 j^z (M^z - 2 O_A) 
\ee
where we have defined:
\be
O_J= \frac{1}{2\pi} \int_{-\pi}^{\pi} dk\, n_k \cos k;
\ee
\be
j^z = -\frac{1}{2 \pi} \int_{-\pi}^{\pi} dk\, n_k \sin k
\label{eq:jz}
\ee
and $O_A$, $O_B$ are given as:
\begin{gather} 
O_B=   \lim_{N\to \infty} \frac{1}{N} \sum_n \langle \left(S_{n-1}^{x}S_{n}^{z}S_{n+1}^{y} \,- S_{n-1}^{y}S_{n}^{z}S_{n+1}^{x}\right)\rangle  \nonumber \\ =  \frac{1}{4\pi} \int_{-\pi}^{\pi} dk \, n_k \sin 2k
\end{gather}
and
\begin{gather}
O_A =  \lim_{N\to \infty} \frac{1}{N} \sum_n \langle \left(S_{n-1}^{x}S_{n}^{z}S_{n+1}^{x}+S_{n-1}^{y}S_{n}^{z}S_{n+1}^{y} \right) \rangle \nonumber \\ = -\frac{1}{4\pi} \int_{-\pi}^{\pi} dk \, n_k \cos 2k 
\end{gather}

\begin{figure}[ht!]
\centering
\includegraphics[width=3.0in]{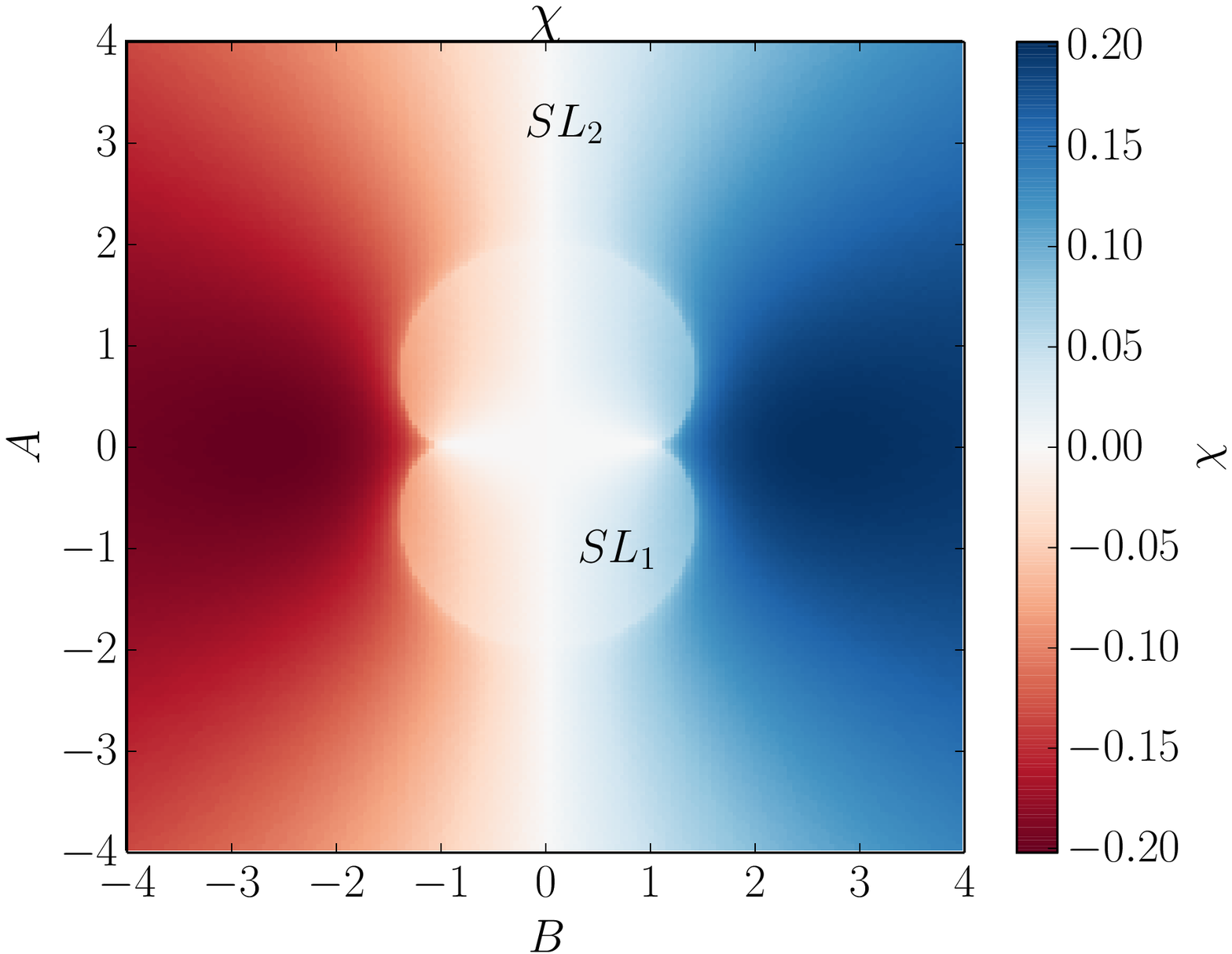}{(a)}
\includegraphics[width=3.0in]{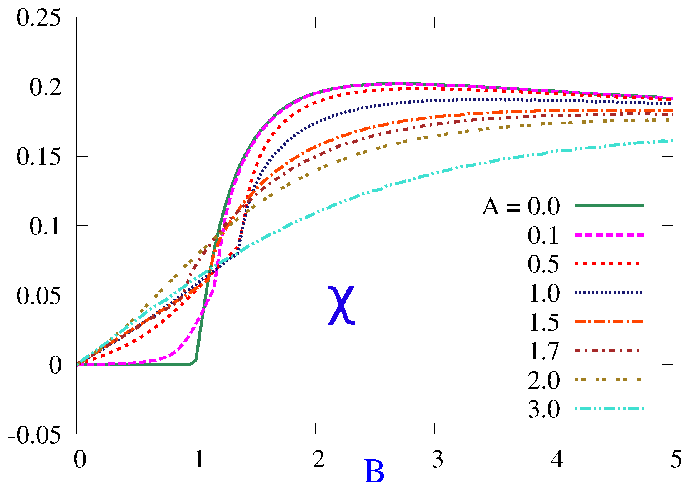}{(b)}
\includegraphics[width=3.0in]{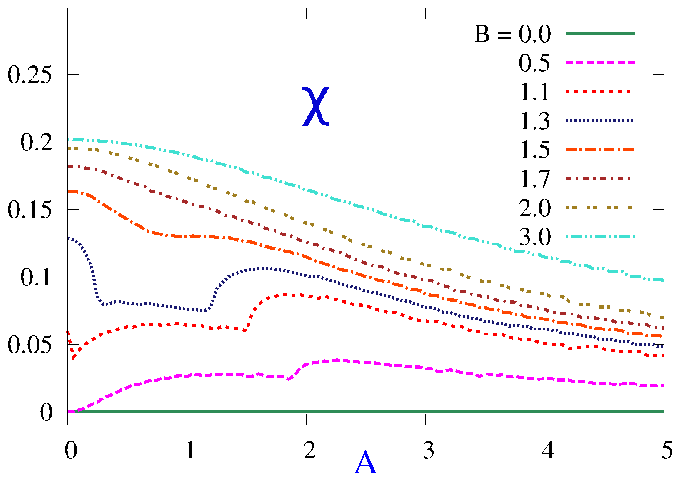}{(c)}
\caption[The $B-A$-dependence of the scalar chirality.]{\label{fig:scalar_chirality} (a) The $B-A$-dependence of the scalar chirality ($T=0.001$). Scalar chirality as a function (b) of $B$ for different values of $A$, and (c) of $A$ for different values of $B$, both at $T=0.01$.} 
\end{figure}

We plot the scalar chirality $\chi$  as a function of $B$ and $A$ in Fig.~\ref{fig:scalar_chirality}.
From Fig.~\ref{fig:scalar_chirality}(a) we can see that the scalar chirality vanishes identically if $B=0$. It can be seen from Fig.~\ref{fig:scalar_chirality}(a, b) that in the absence of $A$, for a finite $B$, $\chi$ vanishes in the $SL_1$ phase, i.e. when $B<B_c$, while for $B>B_c$, there is a finite scalar chirality which saturates to a constant value with increasing $B$. For non-vanishing $A$ and $B$, there is a finite scalar chirality even in the $SL_1$ phase. A finite $A$ value leads to a decrease in the magnitude of the scalar chirality in the $SL_2$ phase as shown in Fig.~\ref{fig:scalar_chirality}(a,b,c). 

We next define the vector chirality operator $\hat{K}_{n,m}$ as:
\be
\hat{K}_{n,m} = \vec S_n \times \vec S_{m}
\ee
The $z$ component of the NN vector chirality order is defined as  $<K_{n,n+1}^z> \equiv \langle \hat{j}^z \rangle = j^z $ (Eq. \ref{eq:jz}).
The electric polarization $P^y$ (Eq.~\ref{eq:p1p2}) is then given as:
\begin{gather}
P^{y } = - \frac{\partial f_{N}(T,A, B, h_z, E)}{\partial E}  = \langle {\hat P}_1^y \rangle  + \langle {\hat P}_2^y \rangle \nonumber
\label{eq:pol}
\end{gather}
where
\begin{gather}
\langle {\hat P}_1^y \rangle = -J j^z; \,\, \langle {\hat P}_2^y \rangle = - 2 A O_B + 2 B O_A
\end{gather}

Substituting the expressions for $j^z$, $O_B$ and $O_A$ in Eq.~\ref{eq:pol}, we have
\begin{gather}
P^y = \frac{1}{2\pi} \int_{-\pi}^{\pi} dk\, n_k \left( J \sin k - A \sin 2k - B \cos 2k \right) \nonumber \\
 = \frac{1}{2\pi} \int_{\Lambda} dk\, \left( J \sin k - A \sin 2k - B \cos 2k \right)
\end{gather} 
Here $\Lambda$ is the momentum region enclosing the occupied states, i.e. where $\varepsilon_k<0$, and is bounded by the Fermi points at which $\varepsilon_k$ vanishes. Since the antiderivative of the integrand is $ - J \cos k + \frac{A}{2} \cos 2k - \frac{B}{2} \sin 2k = -\varepsilon_k$, it vanishes at the limits of integration (the Fermi points). We thus observe that, in the absence of external fields, the total polarization $P^y$ (Eq.~\ref{eq:pol}) is zero for any $B$ and $A$, as also noted in Ref.~\cite{menchyshyn2015}.\\ 

\begin{figure}[ht!]
\includegraphics[width=3.0in]{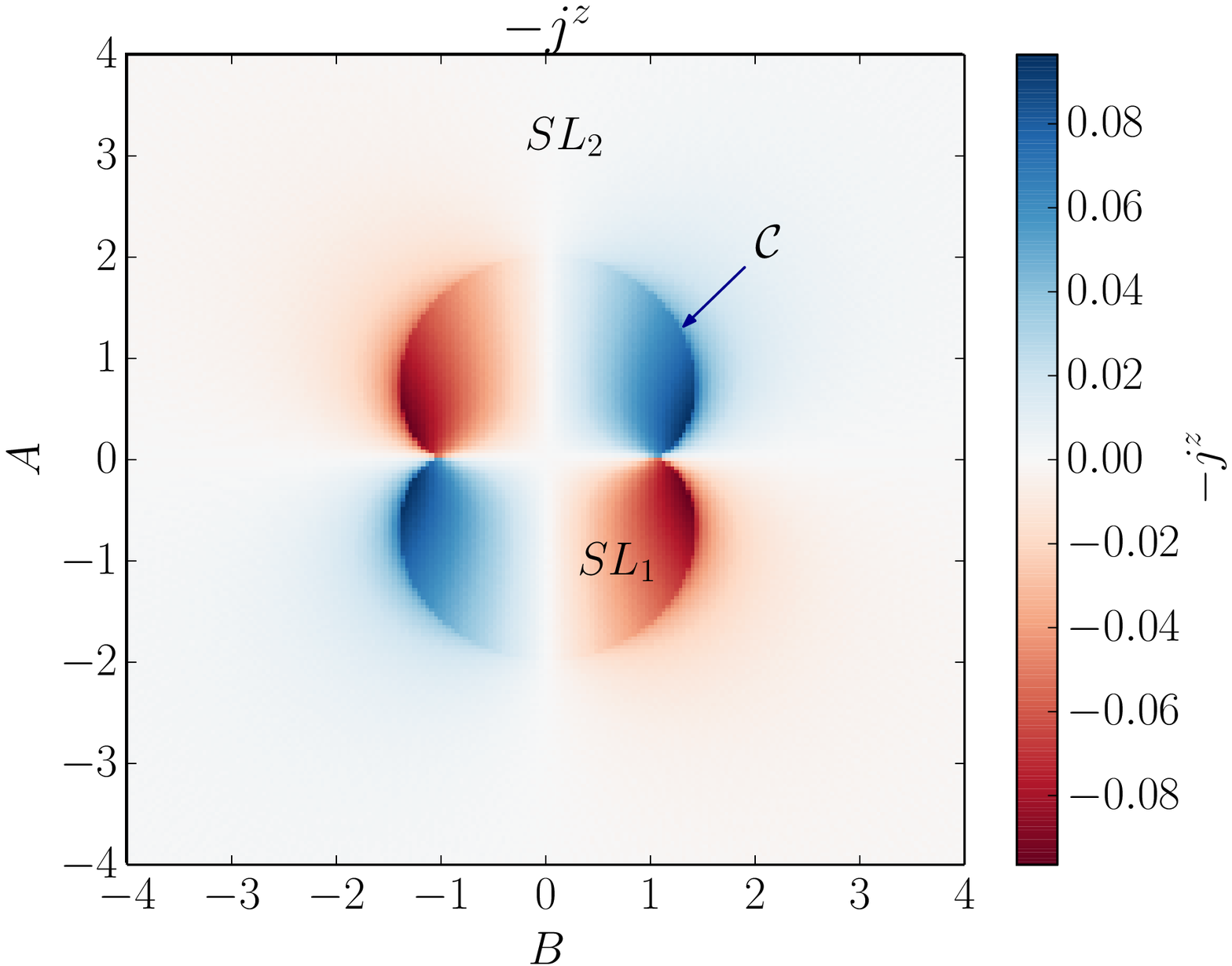}{(a)}
\includegraphics[width=3.0in]{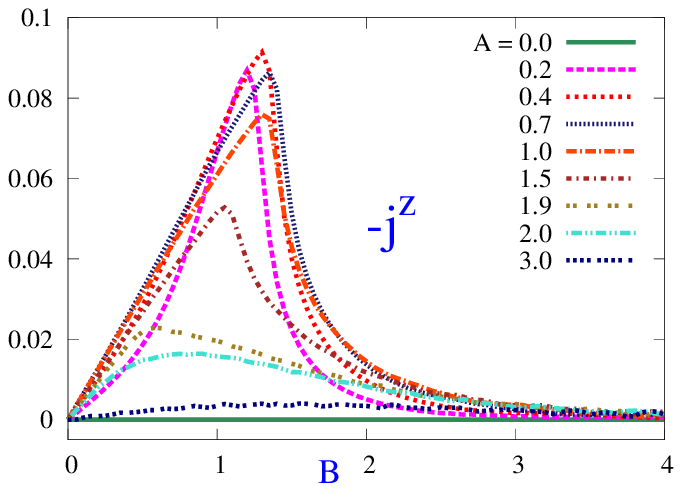}{(b)}
\includegraphics[width=3.0in]{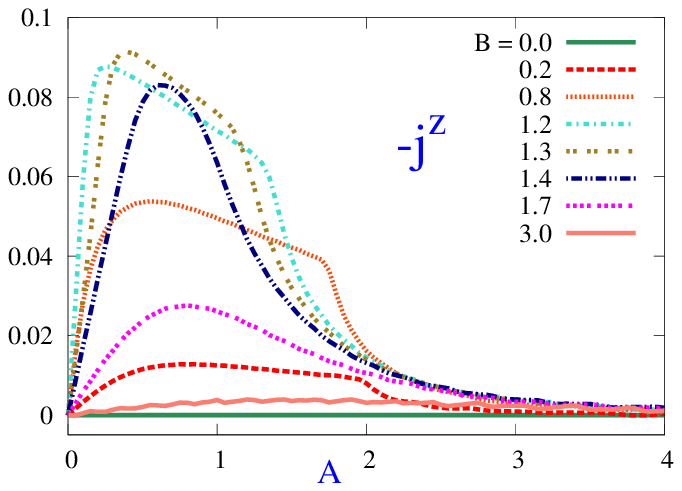}{(c)}
\caption[The $B-A$ dependence of the NN spin current.]{\label{fig:spin_current} (a) The $B-A$ dependence of the NN spin current $j^z$, with $T=0.001$. The spin current as a function of (b) $B$ for different values of $A$ (c) $A$ for different values of $B$, both at $T=0.01$.}
\end{figure}

However, there exists a parameter regime in the $A-B$ plane where $\langle {\hat P}_1^y \rangle$  and $\langle {\hat P}_2^y \rangle$ {\it do not vanish individually}, even though the total polarization $P^y = \langle {\hat P}_1^y \rangle + \langle {\hat P}_2^y \rangle$ itself vanishes.
The non-zero spin current $j^z = \langle {\hat P}_1^y \rangle = -\langle {\hat P}_2^y \rangle$ is related to the existence of a vector chiral order. We describe now the behaviour of the NN spin current $j^z$ as a function of $A$ and $B$. In contrast to the behaviour of the magnetization or the scalar chirality which is finite for non-zero $A(B)$ even when $B(A)$ is zero, $j^z$ vanishes when either $A$ or $B$ is zero. This can be seen from Fig.~\ref{fig:spin_current}(a) where we show the dependence of the NN spin current on the values of $B$ and $A$.
 However, when both $A$ and $B$ are non-zero, there is a finite NN vector chirality for a range of parameters as can be seen from Fig.~\ref{fig:spin_current}.  The spin current shows a non-monotonic $A$ and $B$ dependence, going to a maximum along the critical curve (except when $A=0$ or $B=0$). 
\noindent Again, for large values of $B$ and $A$, the spin current vanishes. The spin current is odd with respect to both $B$ and $A$.
Thus, as seen from the figures above, there exists a parameter regime in the 
$A-B$ plane where, even though $P^y$ itself vanishes, $j^z$ does not vanish. This signifies the existence of a vector chiral order. 

\begin{figure}[ht!]
\includegraphics[width=3.0in]{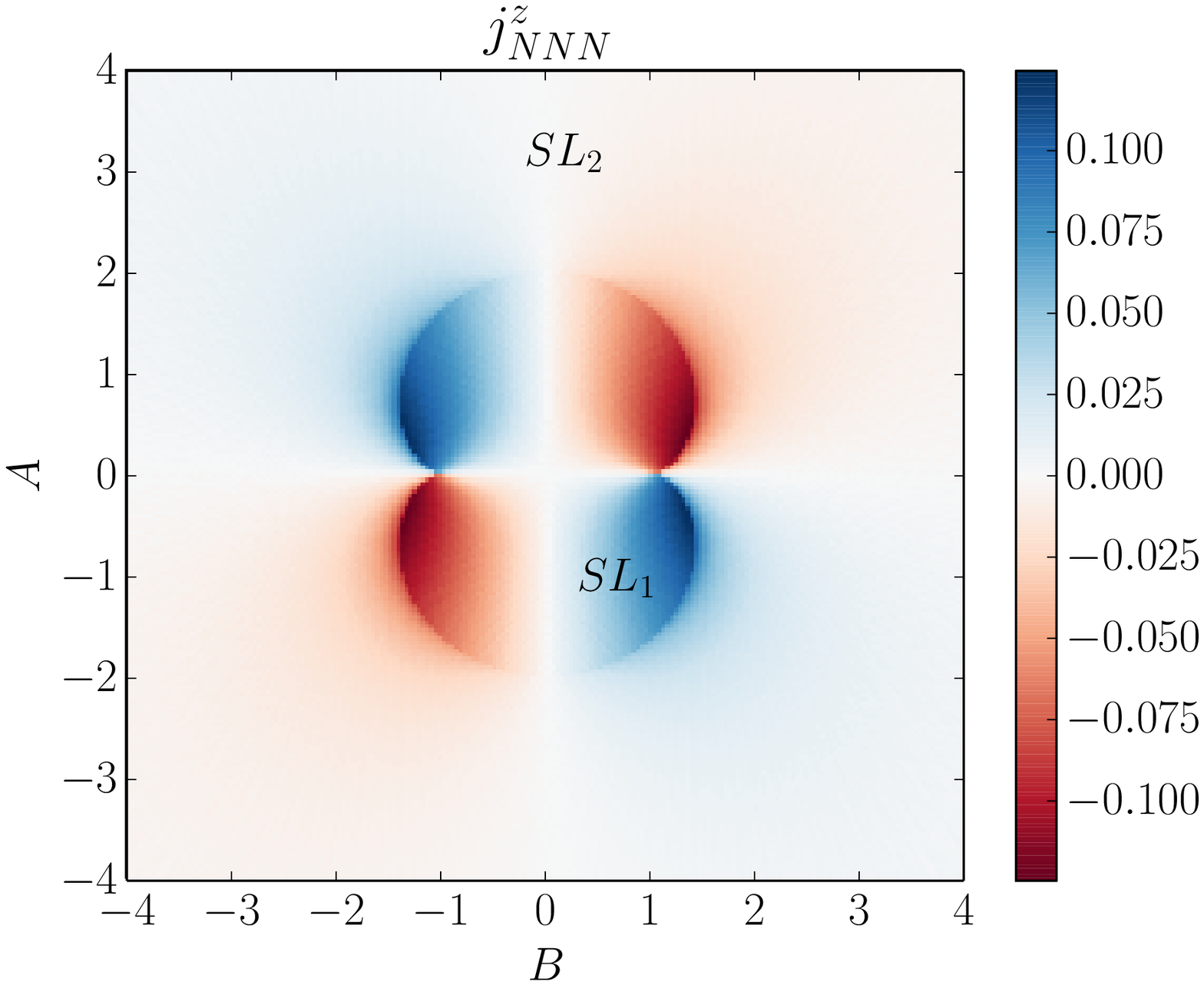}{(a)}
\includegraphics[width=3.0in]{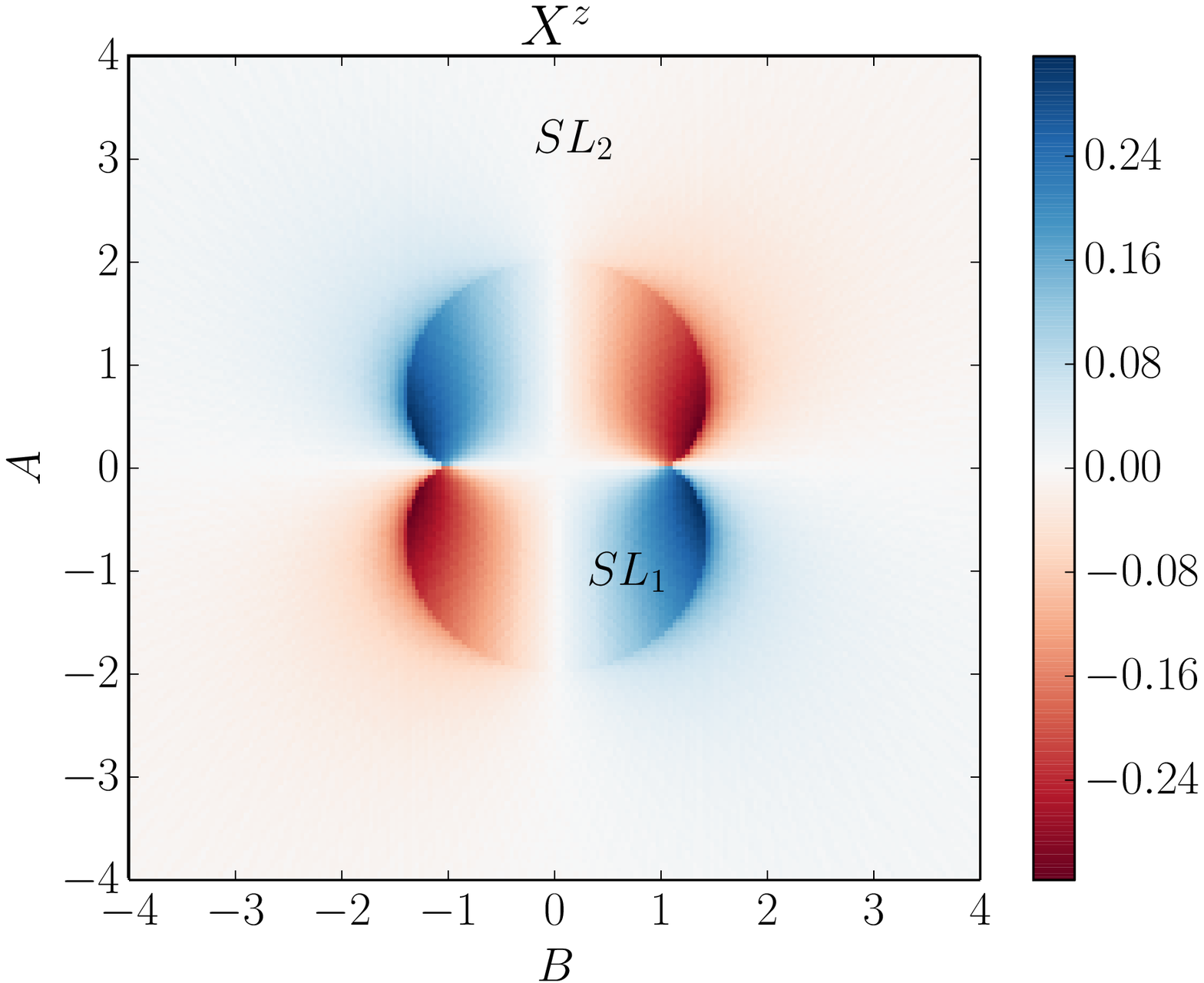}{(b)}
\caption[$BA$-dependence of (a) the NNN spin-current $j^z_{NNN}$, and (b) the plaquette vector chirality $X^z$]{\label{fig:jNNN} $BA$-dependence of (a) the next-nearest neighbour spin-current $j^z_{NNN}$, and (b) the plaquette vector chirality $X^z$, both at $T=0.001$.}
\end{figure}

We also find interesting extended chiral order. 
We define the $z$-component of the next-nearest-neighbour vector chirality $j^z_{NNN}$ as:
\be
j^z_{NNN}\equiv< K_{n+2,n}^z > = -4 [j^z O_J + m^z O_B]
\ee
We show the $B-A$ dependence of  $j^z_{NNN}$ in Fig.~\ref{fig:jNNN}(a).
$j^z_{NNN}$ shows a similar behaviour as $j^z$, in that it vanishes unless both $A$ and $B$ are non-zero as can be seen from the figure. In general, it has a non-monotonic $B-A$ dependence. $j^z_{NNN}$ vanishes as $B$ and $A$ become very large. We note here that such a NNN spin current arises only due to the presence of the higher order three-spin interactions.

\begin{figure}[ht!]
\includegraphics[width=3.0in]{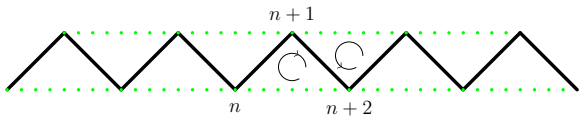}
\caption[Schematic depiction of the triangular plaquettes formed by neighbouring spins.]{\label{fig:plaqs} Schematic depiction of plaquettes and the associated sense of the spin current.}
\end{figure}

We next introduce the plaquette vector chirality as: 
\be
\vec X  =   \lim_{N\to \infty} \frac{1}{N} \sum_n  \langle (\vec K_{n,n+1}  + \vec K_{n+1, n+2} + \vec K_{n+2,n} )\rangle
\ee
The $z$ component of the plaquette vector chirality, $X^z$,  can be easily seen to be given as:
\be
X^z = 2 j^z + j^z_{NNN}
\ee
We  plot the plaquette vector chirality $X^z$ as function of $A$ and $B$ in Fig.~\ref{fig:jNNN}(b).  Again, unlike the behaviour of the magnetization or the scalar chirality which is finite for non-zero 
$A(B)$ even when $B(A)$ is $0$, the plaquette vector chirality  vanishes when either $A$ or $B$ is zero. However, when both $A$ and $B$ are non-zero, then we find that there is a finite plaquette vector chirality for a range of parameters as can be seen 
from the figure. Thus, we find interesting behaviour that when both $A$ and $B$ are present, there is an additional plaquette vector chiral order for a certain parameter regime in the $A-B$ plane. The non-vanishing of the vector chiral order $X^z$ is due to the non-vanishing of both the NN and NNN spin currents. We can therefore identify this regime as a phase with circulating chiral spin currents in the triangular plaquettes $n, n+1, n+2$ as shown in Fig.~\ref{fig:plaqs}. We can also check that the sign of the circulating current is opposite in adjacent plaquettes. 
Such circulating orbital spin currents arise due to the noncollinear ordering of the spins at the lattice sites $n,n+1,n+2$.

 We argue that analogous  to charge current loops giving rise to orbital magnetic dipole moments, the orbital spin currents give rise to orbital electric dipole moments or an electric flux through each plaquette and therefore such a phase gives rise to orbital anti-ferroelectricity (due to opposite signs of the electric flux in adjacent plaquettes) in the one-dimensional system. Thus we find that even in the absence of any external fields, a spontaneous orbital anti-ferroelectricity arises when both $A$ and $B$ are present. The emergence of such a novel orbital electric flux phase is due to the chiral non-coplanar spin structure. It is quantified by both a non-zero plaquette scalar chirality and a persistent chiral spin current loop or the non-zero plaquette vector chirality (which occurs due to a non-zero NN spin current and a non-zero NNN spin current).

    We can understand the existence of this novel phase using the fermionic picture. The presence of two-spin and three-spin interactions imply in the equivalent fermion model, the presence of the NN and NNN hopping with the NNN hopping strength becoming complex (with a non-zero real part) when both types ($XZX+YZY$ and $XZY-YZY$) of three-spin interactions are present. Non-trivial closed loops connecting NN and NNN sites or a non-trivial topological connectivity arises because of the complex higher range interaction (schematically shown in Fig.~\ref{fig:plaqs}). This non-trivial topological connectivity gives rise to a higher dimensional toroidal order. The toroidal order of this novel phase can be characterized by a scalar toroidal order parameter and a vector toroidal order parameter. The scalar toroidal parameter is defined as $ \tau_S=  \chi M^z$; from the behaviour of $\chi$ and $M^z$, we can see that $\tau_S$ is either positive or negative depending on the relative signs of $A$ and $B$. Defining the vector toroidal moment $ \vec{\tau} = \vec{P_1} \times {\vec M}$, we can further characterize the phase by a vector toroidal order parameter $ \tau_V= P_1^y M^z$ along the $x$-direction. Thus, a higher dimensional non-trivial magnetoelectric order emerges even though the model itself is one-dimensional.
Such a connection between higher range interactions and higher dimensionality was noted in Ref.~\cite{grass}.
We do not discuss this here but one can also show that the various orders can be tuned by external electric and magnetic fields as discussed earlier in Refs.~\cite{pradeep2016,pdn2016}. We again mention here that the existence of a non-trivial MEE and the tunability of magnetization by external electric fields in the $JAE$ model was discussed in Ref.~\cite{menchyshyn2015}. We emphasize however that the novel orbital antiferroelectricity in the absence of external fields that we discuss in this paper has not been reported earlier.
In the next section, we describe the non-trivial topological properties of the model by a study of the entanglement entropy.

\section[Global order]{\label{sec:EE}Entanglement entropy and topological order}
In  the conventional approach to the study of QPT, the focus is on studying how local observables change across the QPT and identifying a non-trivial local order associated with the change of symmetry across the QPT. The discovery that distinct phases of matter can occur without change in symmetry has led to a more general notion of order associated with the global topology of the ground state wave functions. The QPT which occurs in the model considered above is one such example of a QPT where no change of symmetry occurs across the QPT. The two phases $SL_1$ and $SL_2$ are both gapless; however they are distinct even though there is no change of symmetry across the QPT. As pointed out earlier, at long wavelengths, the two phases are believed to be described by two CFTs with central charge $c=1$ and $c=2$ respectively. The central charge $c$ is a universal number characterizing the CFT and can be loosely thought of therefore as a non-trivial order parameter characterizing the two phases. 
What clearly changes across the QPT is the number of Fermi points. The change in the Fermi sea topology across the transition makes this a so-called Lifshitz transition\cite{lifshitz1960}. The `connectedness' property of the Fermi sea is the most obvious topological invariant that distinguishes the two phases across the Lifshitz transition. As shown in Sec.~\ref{sec:GSPD}, the $SL_1$ phase corresponds to a phase with two Fermi points and a single `connected' Fermi sea while the $SL_2$ phase corresponds to that with four Fermi points and a `disconnected'  Fermi sea with two disjoint pieces.  The QPT thus ought to be characterised by an order parameter that captures this change in Fermi sea topology\cite{krokhmalskii2008}. 
In the following, we make an attempt to capture the change in the Fermi sea topology by studying the behaviour of two non-local quantities, namely (i) the new occupied $k$-space volume that gets opened up due to the emergence of additional Fermi points and (ii) the entanglement entropy, and probe how these quantities behave as a function of the three-spin coupling strengths $A$ and $B$.

\begin{figure}
\includegraphics[width=3.0in]{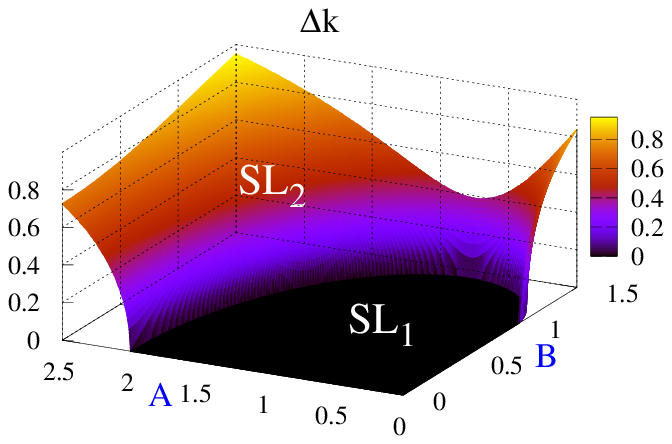}{(a)}
\includegraphics[width=3.0in]{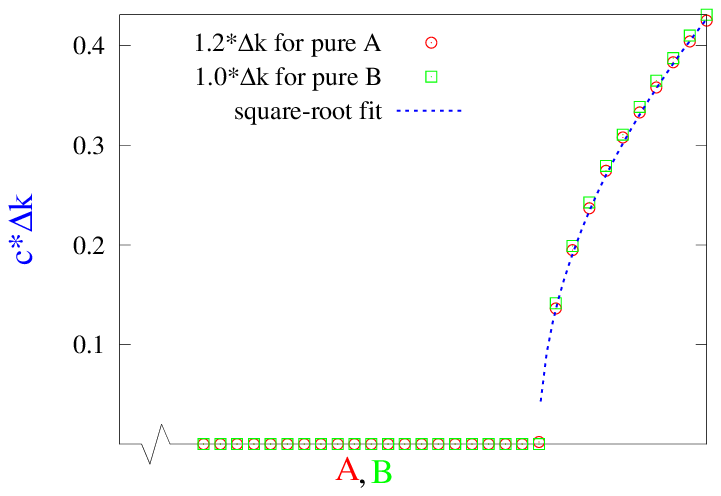}{(b)}
\caption[Increase in size of the new ``piece" of $k$-space, $\Delta k$, beyond the Lifshitz transition in the presence of both $B$ and $A$.]{(a) Increase in size of the new ``piece" of $k$-space, $\Delta k$, beyond the Lifshitz transition in the presence of both $B$ and $A$. The black region in the second panel denotes the situation where the Fermi sea is simply connected. (b) Square-root dependence of $\Delta k$ on the three-spin interaction strength.}
\label{fig:kspaceinc}
\end{figure}

 We begin by studying the behaviour of the new occupied region (volume of $k$-space) which emerges in the $SL_2$ phase due to the additional Fermi points.  In Fig.~\ref{fig:kspaceinc}(a), we plot the $B-A$ dependence of size (length) of the new piece $\Delta k$ of the Fermi sea.
$\Delta k$ is zero in the $SL_1$ phase (the black region). It becomes finite as we cross into the $SL_2$ phase. Near the QCP, $\Delta k$ shows a square root dependence on the three-spin coupling strengths. The non-analytic square root behaviour can be see explicitly in panel (b) where we plot $\Delta k$ for the cases $B=0$ and $A=0$. (In the latter case, we have plotted $\Delta k$ scaled suitably.)
$\Delta k$ saturates to the value $\pi /2$ for very large $A,B$. 

   Another non-trivial quantity which has been used as a diagnostic to detect non-trivial topological order in gapless phases is the the entanglement entropy (EE). The properties of entanglement entropy  in one-dimensional conformal field theories (CFT) has been studied for a very long time now. The EE of a one-dimensional CFT  with central charge $c$ has the feature of universal scaling behaviour.
For a system of non-interacting fermions, the entanglement entropy of a block of fermions in a region of size $l$ is given as:
\begin{equation}
S(\textit{l}) = -\text{Tr} \left[ M \text{ln} M + (1-M) \text{ln}  (1-M) \right]
\end{equation}
where the matrix of fermionic correlations, $M$ is defined as \cite{peschel2003}:
\begin{equation}
M_{ij}(\textit{l}) = \langle c_i^{\dagger} c_j\rangle = \frac{1}{2\pi}\int_{-\pi}^{\pi} dk \hspace*{2mm} n_k e^{-ik(i-j)}
\end{equation}
Here $\mathit{l}$ is the length of the subsystem (the number of sites) selected from inside the long chain.
In the absence of three-spin interactions, the EE for long subsystem sizes $\textit{l} \rightarrow \infty$ has been shown to be of the form\cite{jin2004}:
\begin{equation}
S(\textit{l}) \sim  \frac{m}{3}\text{ln}(\textit{l})+ \text{constant}
\end{equation} 
where $m$, the prefactor of the leading logarithmic term is half the number of Fermi points and determines the central charge of the underlying CFT.

\begin{figure}[!htpb]
\includegraphics[width=3.0in]{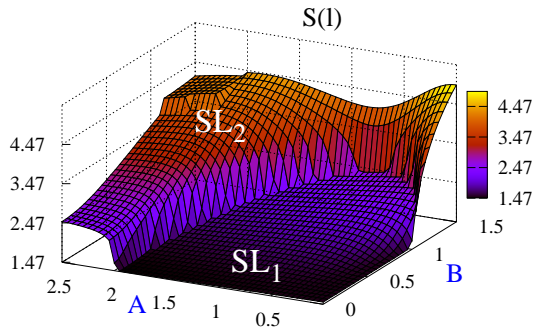}
\caption[The $B-A$-dependence of entanglement entropy.]{The $B-A$-dependence of entanglement entropy for a subsystem of $11$ sites.}
\label{fig:Sl_KE_2d}
\end{figure}
In Fig.~\ref{fig:Sl_KE_2d}, we show the $A-B$-dependence of $S(\textit{l})$. A dramatic rise in $S(\textit{l})$ is observed inside the $SL_2$ phase, in the vicinity of the critical curve that demarcates the transition between phases $SL_1$ and $SL_2$. 
The rate of rise in the $SL_1$ and $SL_2$ phases is different. The change in $S(\textit{l})$ with three-spin coupling strength is more or less regular in the $SL_1$ phase, while in the $SL_2$ phase, the EE rises and saturates for very large coupling strengths. 

\begin{figure*}[!htpb]
\includegraphics[width=3.0in]{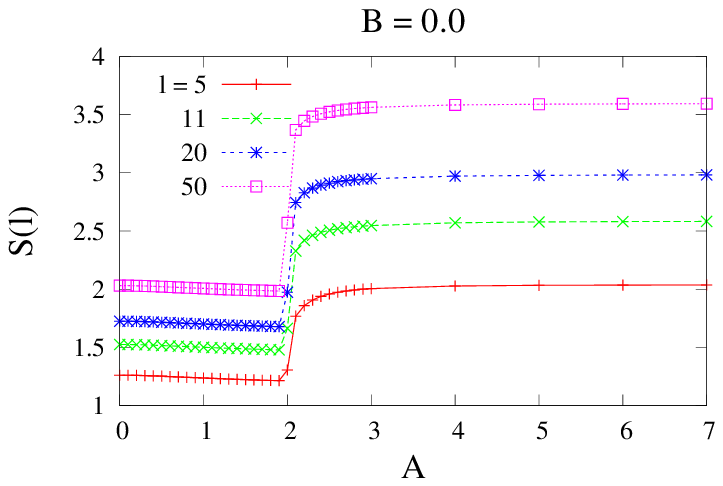}{(a)}
\includegraphics[width=3.0in]{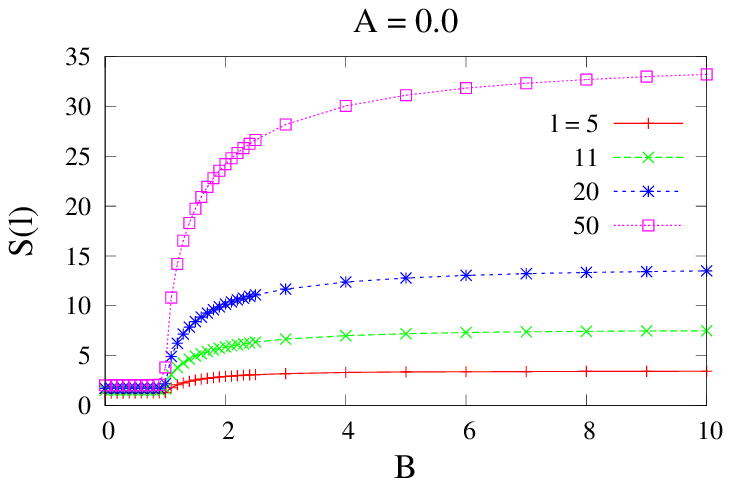}{(b)}
\includegraphics[width=3.0in]{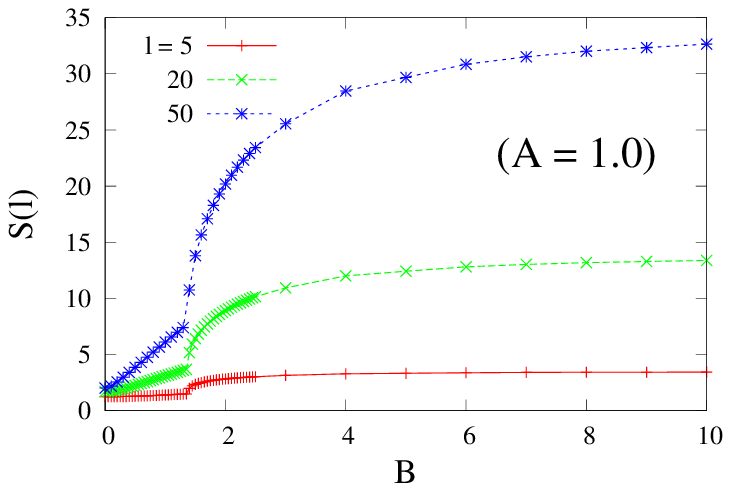}{(c)}
\includegraphics[width=3.0in]{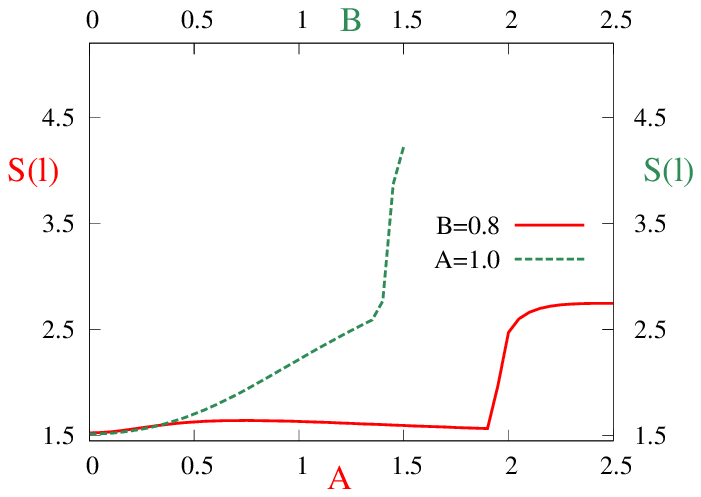}{(d)}
\caption[$A$- and $B$-dependence of the entanglement entropy.]{Dependence of EE (a) on $B\,(A=0)$, (b) on $A\,(B=0)$, and (c) on $B\,(A=1.0)$), all plotted for different lengths. Abrupt jumps are observed at the critical points marking the Lifshitz transition. (d) $A$- and $B$-dependence of entanglement entropy for subsystem size $\textit{l}=11$, in the presence of the other $3$-spin coupling. There is a nontrivial $B$-dependence of $S(\textit{l})$ in the $SL_1$ phase in the presence of $A$.}
\label{fig:Sl_1d}
\end{figure*}

This can be seen better from Fig.~\ref{fig:Sl_1d} where we show the dependence of the EE on the three-spin couplings for different lengths.

The magnitude of the EE changes with length, but some common qualitative features are observed. For the pure $A$ case (Fig.~\ref{fig:Sl_1d}(a)), the EE decreases for $A < A_c \approx 2$, with very small but constant slope, shows an abrupt jump at $A = A_c$ and then saturates to a constant value for large $A$ (different for different lengths). The abrupt jump marks the Lifshitz transition~\cite{rodney2013}, separating the phases with different Fermi sea topologies.
For the pure $B$ case shown in Fig.~\ref{fig:Sl_1d}(b), it can be seen that the magnitude of the EE shows a more marked length dependence compared to that for the pure $A$ case.
But again, some common features are observed. For $B < B_c \approx 1.0$, EE is a length-dependent constant (although not seen very clearly in the figure shown due to the scale), independent of $B$ unlike that for the pure $A$ case where there was a small negative slope. At $B = B_c$, there is a jump in the EE indicating the Lifshitz transition. 
For large values of $B$ beyond $B_c$, the EE becomes flatter and tends towards a saturation value.

In Fig.~\ref{fig:Sl_1d}(c), we plot the $B$-dependence of EE for a nonzero fixed value of $A$ for different subsystem lengths $\textit{l}=5,20,50$.  
The EE again shows abrupt change in the value at the critical coupling value where the QPT occurs.
It can be seen from the plot that  $S(\textit{l})$ shows a nontrivial $B$-dependence inside the $SL_1$ phase unlike that for the pure $A$ or pure $B$ cases.
One finds that for $B < B_c$, the EE now shows an approximately linear increase with $B$, with the rate of increase increasing with subsystem length. For $B>B_c$ the behaviour of the EE is qualitatively similar to that for the pure $B$ case. In Fig.~\ref{fig:Sl_1d}(d), we plot the $A$-dependence of EE for a nonzero fixed value of $B$. Again, one can see that the EE has a non-trivial non-linear $A$ dependence in the $SL_1$ phase.

We mention here that that the EE for the case of a pure three-spin interaction of the $XZX +YZY$ type was computed and shown to fit to a form\cite{frahm1992, rodney2013}:
\begin{equation}
S(\textit{l}) = \alpha(A)\frac{1}{3}\text{ln}(\textit{l})+b(A)
\end{equation}    
In particular, it was shown that for large subsystem sizes, the prefactor $\alpha(A)$ jumps from the value $1$ to the value $2$ across the QCP indicating the change in the central charge from $c=1$ to $c=2$ across the Lifshitz transition.

\section{Conclusion}\label{sec:conc}

In conclusion, we have studied the spin-$1/2$ $XX$ model extended by the three-spin interactions of the $XZX+YZY$ and $XZY-YZX$ types and described its exotic physical properties when both kinds of three spin interactions are taken into account. We solved the model exactly and obtained the ground state phase diagram as a function of the two three-spin coupling strengths. 
An intriguing observation that we made in this context is that the critical curve corresponding to the maximal entropy curve is not smooth at the points where the global maximum of the entropy occurs; the curvature diverges at these points. We showed that \textit{even in the absence of external electric and magnetic fields} there is a phase which exhibits spontaneous magneto-electric order when both $XZX+YZY$ and $XZY-YZX$ types of interaction are present. Specifically, in this regime, we showed the existence of a non-zero magnetization as well as scalar and vector chiral orders. Further, we showed the existence of a plaquette vector chirality  or in other words circulating chiral spin current loops in the plaquettes $n, n+1, n+2$ with the sense of the current being opposite in adjacent plaquettes. Analogous to charge current loops giving rise to orbital magnetic dipole moments, the circulating spin current loops give rise to orbital electric dipole moments. This gives rise to a novel orbital antiferroelectricity. We characterize this phase by a scalar and vector toroidal order.
This new phase with higher dimensional order arises because of the non-trivial topological connectivity resulting from the presence of both the three-spin interactions. We also studied the non-trivial topological properties by studying the the entanglement properties of the model. We observe an interesting nontrivial dependence of the entanglement entropy on the strength of the $XZY-YZX$ interaction in the presence of $XZX+YZY$ interaction.

 We also mention here that there is a lot of experimental interest in the study of models with higher-order multilinear spin interactions~\cite{grytsiuk}.
 Three-spin interactions have been shown to be experimentally accessible in cold atom set-ups coupled with optical lattices~\cite{pachos}. Appropriate tuning of the couplings in such set-ups allows for the free variation of the multi-spin terms in the corresponding Hamiltonian. The nontrivial topological connectivity that arises in our model due to higher-order spin interactions is an example of synthetic dimensionality that can be realized in cold atom experiments which use ultrafast periodic driving ~\cite{grass, ozawa2019}. Recently, it has been shown that a three-spin scalar-chiral interaction, which can be either topology-induced or induced by spin-orbit coupling, can by itself lead to stabilization of noncollinear magnetic textures in some materials~\cite{mankovsky}. 
We expect our study to prove useful in these contexts.

\backmatter

\bmhead{Acknowledgments}

PT acknowledges his discussion with Hemlata Bhandari on entanglement entropy. PT thanks the University of Pune for doctoral stipend and the University Grants Commission, India, for awarding the Basic Scientific Research Fellowship 2015. PD thanks SERB, India, for financial support through research grant MTR/2019/001411.

\section*{Declarations}

\bmhead{Data availability statement} This manuscript has no associated data.

\bmhead{Author contributions}
Both authors contributed equally to this paper.

\bibliography{arXiv2023}


\begin{thebibliography}{34}
\ifx \bisbn   \undefined \def \bisbn  #1{ISBN #1}\fi
\ifx \binits  \undefined \def \binits#1{#1}\fi
\ifx \bauthor  \undefined \def \bauthor#1{#1}\fi
\ifx \batitle  \undefined \def \batitle#1{#1}\fi
\ifx \bjtitle  \undefined \def \bjtitle#1{#1}\fi
\ifx \bvolume  \undefined \def \bvolume#1{\textbf{#1}}\fi
\ifx \byear  \undefined \def \byear#1{#1}\fi
\ifx \bissue  \undefined \def \bissue#1{#1}\fi
\ifx \bfpage  \undefined \def \bfpage#1{#1}\fi
\ifx \blpage  \undefined \def \blpage #1{#1}\fi
\ifx \burl  \undefined \def \burl#1{\textsf{#1}}\fi
\ifx \doiurl  \undefined \def \doiurl#1{\url{https://doi.org/#1}}\fi
\ifx \betal  \undefined \def \betal{\textit{et al.}}\fi
\ifx \binstitute  \undefined \def \binstitute#1{#1}\fi
\ifx \binstitutionaled  \undefined \def \binstitutionaled#1{#1}\fi
\ifx \bctitle  \undefined \def \bctitle#1{#1}\fi
\ifx \beditor  \undefined \def \beditor#1{#1}\fi
\ifx \bpublisher  \undefined \def \bpublisher#1{#1}\fi
\ifx \bbtitle  \undefined \def \bbtitle#1{#1}\fi
\ifx \bedition  \undefined \def \bedition#1{#1}\fi
\ifx \bseriesno  \undefined \def \bseriesno#1{#1}\fi
\ifx \blocation  \undefined \def \blocation#1{#1}\fi
\ifx \bsertitle  \undefined \def \bsertitle#1{#1}\fi
\ifx \bsnm \undefined \def \bsnm#1{#1}\fi
\ifx \bsuffix \undefined \def \bsuffix#1{#1}\fi
\ifx \bparticle \undefined \def \bparticle#1{#1}\fi
\ifx \barticle \undefined \def \barticle#1{#1}\fi
\bibcommenthead
\ifx \bconfdate \undefined \def \bconfdate #1{#1}\fi
\ifx \botherref \undefined \def \botherref #1{#1}\fi
\ifx \url \undefined \def \url#1{\textsf{#1}}\fi
\ifx \bchapter \undefined \def \bchapter#1{#1}\fi
\ifx \bbook \undefined \def \bbook#1{#1}\fi
\ifx \bcomment \undefined \def \bcomment#1{#1}\fi
\ifx \oauthor \undefined \def \oauthor#1{#1}\fi
\ifx \citeauthoryear \undefined \def \citeauthoryear#1{#1}\fi
\ifx \endbibitem  \undefined \def \endbibitem {}\fi
\ifx \bconflocation  \undefined \def \bconflocation#1{#1}\fi
\ifx \arxivurl  \undefined \def \arxivurl#1{\textsf{#1}}\fi
\csname PreBibitemsHook\endcsname

%%% 1
\bibitem{mf-review}
\begin{barticle}
\bauthor{\bsnm{Tokura}, \binits{Y.}},
\bauthor{\bsnm{Seki}, \binits{S.}},
\bauthor{\bsnm{Nagaosa}, \binits{N.}}:
\batitle{Multiferroics of spin origin}.
\bjtitle{Reports on Progress in Physics}
\bvolume{77}(\bissue{7}),
\bfpage{076501}
(\byear{2014})
\end{barticle}
\endbibitem

%%% 2
\bibitem{curieMEE}
\begin{barticle}
\bauthor{\bsnm{Curie}, \binits{P.}}:
\batitle{First consideration of an intrinsic correlation of magnetic and
  electric properties in a solid}.
\bjtitle{J. de Physique (3rd series)}
\bvolume{3},
\bfpage{393}--\blpage{415}
(\byear{1894})
\end{barticle}
\endbibitem

%%% 3
\bibitem{fiebig2005}
\begin{barticle}
\bauthor{\bsnm{Fiebig}, \binits{M.}}:
\batitle{Revival of the magnetoelectric effect}.
\bjtitle{Journal of physics D: applied physics}
\bvolume{38}(\bissue{8}),
\bfpage{123}
(\byear{2005})
\end{barticle}
\endbibitem

%%% 4
\bibitem{fina2017}
\begin{barticle}
\bauthor{\bsnm{{Fina}}, \binits{I.}},
\bauthor{\bsnm{{Marti}}, \binits{X.}}:
\batitle{Electric {C}ontrol of {A}ntiferromagnets}.
\bjtitle{IEEE Transactions on Magnetics}
\bvolume{53}(\bissue{2}),
\bfpage{1}--\blpage{7}
(\byear{2017}).
\doiurl{10.1109/TMAG.2016.2606561}
\end{barticle}
\endbibitem

%%% 5
\bibitem{noda2007}
\begin{barticle}
\bauthor{\bsnm{Noda}, \binits{K.}},
\bauthor{\bsnm{Akaki}, \binits{M.}},
\bauthor{\bsnm{Nakamura}, \binits{F.}},
\bauthor{\bsnm{Akahoshi}, \binits{D.}},
\bauthor{\bsnm{Kuwahara}, \binits{H.}}:
\batitle{Internal magnetic field effect on magnetoelectricity in orthorhombic
  {RM}n{O}$_3$ crystals}.
\bjtitle{Journal of Magnetism and Magnetic Materials}
\bvolume{310}(\bissue{2, Part 2}),
\bfpage{1162}--\blpage{1164}
(\byear{2007}).
\doiurl{10.1016/j.jmmm.2006.10.337}.
\bcomment{Proceedings of the 17th International Conference on Magnetism}
\end{barticle}
\endbibitem

%%% 6
\bibitem{mikeska-review}
\begin{bchapter}
\bauthor{\bsnm{Mikeska}, \binits{H.-J.}},
\bauthor{\bsnm{Kolezhuk}, \binits{A.K.}}:
\bctitle{One-dimensional magnetism}.
In: \beditor{\bsnm{Schollw{\"o}ck}, \binits{U.}},
\beditor{\bsnm{Richter}, \binits{J.}},
\beditor{\bsnm{Farnell}, \binits{D.J.J.}},
\beditor{\bsnm{Bishop}, \binits{R.F.}} (eds.)
\bbtitle{Quantum Magnetism},
pp. \bfpage{1}--\blpage{83}.
\bpublisher{Springer},
\blocation{Berlin, Heidelberg}
(\byear{2004}).
\doiurl{10.1007/BFb0119591}.
\burl{http://dx.doi.org/10.1007/BFb0119591}
\end{bchapter}
\endbibitem

%%% 7
\bibitem{DM1}
\begin{barticle}
\bauthor{\bsnm{Dzyaloshinsky}, \binits{I.}}:
\batitle{A thermodynamic theory of “weak” ferromagnetism of
  antiferromagnetics}.
\bjtitle{Journal of Physics and Chemistry of Solids}
\bvolume{4}(\bissue{4}),
\bfpage{241}--\blpage{255}
(\byear{1958}).
\doiurl{10.1016/0022-3697(58)90076-3}
\end{barticle}
\endbibitem

%%% 8
\bibitem{DM2}
\begin{barticle}
\bauthor{\bsnm{Moriya}, \binits{T.}}:
\batitle{Anisotropic {S}uperexchange {I}nteraction and {W}eak
  {F}erromagnetism}.
\bjtitle{Phys. Rev.}
\bvolume{120},
\bfpage{91}--\blpage{98}
(\byear{1960}).
\doiurl{10.1103/PhysRev.120.91}
\end{barticle}
\endbibitem

%%% 9
\bibitem{knb}
\begin{barticle}
\bauthor{\bsnm{Katsura}, \binits{H.}},
\bauthor{\bsnm{Nagaosa}, \binits{N.}},
\bauthor{\bsnm{Balatsky}, \binits{A.V.}}:
\batitle{Spin {C}urrent and {M}agnetoelectric {E}ffect in {N}oncollinear
  {M}agnets}.
\bjtitle{Phys. Rev. Lett.}
\bvolume{95},
\bfpage{057205}
(\byear{2005}).
\doiurl{10.1103/PhysRevLett.95.057205}
\end{barticle}
\endbibitem

%%% 10
\bibitem{pradeepPRB1}
\begin{barticle}
\bauthor{\bsnm{Thakur}, \binits{P.}},
\bauthor{\bsnm{Durganandini}, \binits{P.}}:
\batitle{Heisenberg spin-$\frac{1}{2}$ $\mathrm{XXZ}$ chain in the presence of
  electric and magnetic fields}.
\bjtitle{Phys. Rev. B}
\bvolume{97},
\bfpage{064413}
(\byear{2018})
\end{barticle}
\endbibitem

%%% 11
\bibitem{pradeepPRB2}
\begin{barticle}
\bauthor{\bsnm{Thakur}, \binits{P.}},
\bauthor{\bsnm{Durganandini}, \binits{P.}}:
\batitle{Factorization, coherence, and asymmetry in the {H}eisenberg
  spin-$\frac{1}{2}$ $\mathrm{XXZ}$ chain with {D}zyaloshinskii-{M}oriya
  interaction in transverse magnetic field}.
\bjtitle{Phys. Rev. B}
\bvolume{102},
\bfpage{064409}
(\byear{2020}).
\doiurl{10.1103/PhysRevB.102.064409}
\end{barticle}
\endbibitem

%%% 12
\bibitem{pradeep2016}
\begin{barticle}
\bauthor{\bsnm{Thakur}, \binits{P.}},
\bauthor{\bsnm{Durganandini}, \binits{P.}}:
\batitle{Magnetic and electric order in the spin-1/2 {XX} model with three-spin
  interactions}.
\bjtitle{AIP Conference Proceedings}
\bvolume{1731}(\bissue{1}),
\bfpage{130051}
(\byear{2016}).
\doiurl{10.1063/1.4948157}
\end{barticle}
\endbibitem

%%% 13
\bibitem{pdn2016}
\begin{bchapter}
\bauthor{\bsnm{Durganandini}, \binits{P.}}:
\bctitle{Magnetoelectric effects in the spin-$1/2$ {XX} chain with three spin
  interactions and {D}zyaloshinskii-{M}oriya interaction}.
In: \bbtitle{APS March Meeting Abstracts},
vol. \bseriesno{2016},
pp. \bfpage{6}--\blpage{008}
(\byear{2016})
\end{bchapter}
\endbibitem

%%% 14
\bibitem{suzukiPLA1971}
\begin{barticle}
\bauthor{\bsnm{Suzuki}, \binits{M.}}:
\batitle{Equivalence of the two-dimensional {I}sing model to the ground state
  of the linear $\mathrm{XY}$-model}.
\bjtitle{Physics Letters A}
\bvolume{34}(\bissue{2}),
\bfpage{94}--\blpage{95}
(\byear{1971})
\end{barticle}
\endbibitem

%%% 15
\bibitem{suzukiPTP1971}
\begin{barticle}
\bauthor{\bsnm{Suzuki}, \binits{M.}}:
\batitle{Relationship among {E}xactly {S}oluble {M}odels of {C}ritical
  {P}henomena. i*)$2${D} {I}sing {M}odel, {D}imer {P}roblem and the
  {G}eneralized $\mathrm{XY}$-model}.
\bjtitle{Progress of Theoretical Physics}
\bvolume{46}(\bissue{5}),
\bfpage{1337}--\blpage{1359}
(\byear{1971}).
\doiurl{10.1143/PTP.46.1337}
\end{barticle}
\endbibitem

%%% 16
\bibitem{baxterwu}
\begin{barticle}
\bauthor{\bsnm{Baxter}, \binits{R.}},
\bauthor{\bsnm{Wu}, \binits{F.}}:
\batitle{Exact solution of an {I}sing model with three-spin interactions on a
  triangular lattice}.
\bjtitle{Physical Review Letters}
\bvolume{31}(\bissue{21}),
\bfpage{1294}
(\byear{1973})
\end{barticle}
\endbibitem

%%% 17
\bibitem{penson1982}
\begin{barticle}
\bauthor{\bsnm{Penson}, \binits{K.}},
\bauthor{\bsnm{Jullien}, \binits{R.}},
\bauthor{\bsnm{Pfeuty}, \binits{P.}}:
\batitle{Phase transitions in systems with multispin interactions}.
\bjtitle{Physical Review B}
\bvolume{26}(\bissue{11}),
\bfpage{6334}
(\byear{1982})
\end{barticle}
\endbibitem

%%% 18
\bibitem{penson1988}
\begin{barticle}
\bauthor{\bsnm{Penson}, \binits{K.}},
\bauthor{\bsnm{Debierre}, \binits{J.}},
\bauthor{\bsnm{Turban}, \binits{L.}}:
\batitle{Conformal invariance and critical behavior of a quantum {H}amiltonian
  with three-spin coupling in a longitudinal field}.
\bjtitle{Physical Review B}
\bvolume{37}(\bissue{13}),
\bfpage{7884}
(\byear{1988})
\end{barticle}
\endbibitem

%%% 19
\bibitem{gottlieb}
\begin{barticle}
\bauthor{\bsnm{Gottlieb}, \binits{D.}},
\bauthor{\bsnm{R{\"o}ssler}, \binits{J.}}:
\batitle{Exact solution of a spin chain with binary and ternary interactions of
  dzialoshinsky-moriya type}.
\bjtitle{Physical Review B}
\bvolume{60}(\bissue{13}),
\bfpage{9232}
(\byear{1999})
\end{barticle}
\endbibitem

%%% 20
\bibitem{titvinidze2003}
\begin{barticle}
\bauthor{\bsnm{Titvinidze}, \binits{I.}},
\bauthor{\bsnm{Japaridze}, \binits{G.I.}}:
\batitle{Phase diagram of the spin-$1/2$ extended {XY} model}.
\bjtitle{The European Physical Journal B-Condensed Matter and Complex Systems}
\bvolume{32}(\bissue{3}),
\bfpage{383}--\blpage{393}
(\byear{2003})
\end{barticle}
\endbibitem

%%% 21
\bibitem{lou2004}
\begin{barticle}
\bauthor{\bsnm{Lou}, \binits{P.}},
\bauthor{\bsnm{Wu}, \binits{W.-C.}},
\bauthor{\bsnm{Chang}, \binits{M.-C.}}:
\batitle{Quantum phase transition in spin-$1/2$ {XX} {H}eisenberg chain with
  three-spin interaction}.
\bjtitle{Phys. Rev. B}
\bvolume{70}(\bissue{6}),
\bfpage{064405}
(\byear{2004})
\end{barticle}
\endbibitem

%%% 22
\bibitem{krokhmalskii2008}
\begin{barticle}
\bauthor{\bsnm{Krokhmalskii}, \binits{T.}},
\bauthor{\bsnm{Derzhko}, \binits{O.}},
\bauthor{\bsnm{Stolze}, \binits{J.}},
\bauthor{\bsnm{Verkholyak}, \binits{T.}}:
\batitle{Dynamic properties of the spin-$\frac{1}{2}$ {XY} chain with
  three-site interactions}.
\bjtitle{Phys. Rev. B}
\bvolume{77},
\bfpage{174404}
(\byear{2008}).
\doiurl{10.1103/PhysRevB.77.174404}
\end{barticle}
\endbibitem

%%% 23
\bibitem{topilko2012}
\begin{barticle}
\bauthor{\bsnm{Topilko}, \binits{M.}},
\bauthor{\bsnm{Krokhmalskii}, \binits{T.}},
\bauthor{\bsnm{Derzhko}, \binits{O.}},
\bauthor{\bsnm{Ohanyan}, \binits{V.}}:
\batitle{Magnetocaloric effect in spin-1/2 {XX} chains with three-spin
  interactions}.
\bjtitle{The European Physical Journal B}
\bvolume{85}(\bissue{8}),
\bfpage{278}
(\byear{2012})
\end{barticle}
\endbibitem

%%% 24
\bibitem{menchyshyn2015}
\begin{barticle}
\bauthor{\bsnm{Menchyshyn}, \binits{O.}},
\bauthor{\bsnm{Ohanyan}, \binits{V.}},
\bauthor{\bsnm{Verkholyak}, \binits{T.}},
\bauthor{\bsnm{Krokhmalskii}, \binits{T.}},
\bauthor{\bsnm{Derzhko}, \binits{O.}}:
\batitle{Magnetism-driven ferroelectricity in spin-$\frac{1}{2}$ {XY} chains}.
\bjtitle{Phys. Rev. B}
\bvolume{92},
\bfpage{184427}
(\byear{2015})
\end{barticle}
\endbibitem

%%% 25
\bibitem{lifshitz1960}
\begin{barticle}
\bauthor{\bsnm{Lifshitz}, \binits{I.}}, \betal:
\batitle{Anomalies of electron characteristics of a metal in the high pressure
  region}.
\bjtitle{Sov. Phys. JETP}
\bvolume{11}(\bissue{5}),
\bfpage{1130}--\blpage{1135}
(\byear{1960})
\end{barticle}
\endbibitem

%%% 26
\bibitem{grass}
\begin{barticle}
\bauthor{\bsnm{Gra\ss{}}, \binits{T.}},
\bauthor{\bsnm{Muschik}, \binits{C.}},
\bauthor{\bsnm{Celi}, \binits{A.}},
\bauthor{\bsnm{Chhajlany}, \binits{R.W.}},
\bauthor{\bsnm{Lewenstein}, \binits{M.}}:
\batitle{Synthetic magnetic fluxes and topological order in one-dimensional
  spin systems}.
\bjtitle{Phys. Rev. A}
\bvolume{91},
\bfpage{063612}
(\byear{2015}).
\doiurl{10.1103/PhysRevA.91.063612}
\end{barticle}
\endbibitem

%%% 27
\bibitem{peschel2003}
\begin{barticle}
\bauthor{\bsnm{Peschel}, \binits{I.}}:
\batitle{Calculation of reduced density matrices from correlation functions}.
\bjtitle{Journal of Physics A: Mathematical and General}
\bvolume{36}(\bissue{14}),
\bfpage{205}--\blpage{208}
(\byear{2003}).
\doiurl{10.1088/0305-4470/36/14/101}
\end{barticle}
\endbibitem

%%% 28
\bibitem{jin2004}
\begin{barticle}
\bauthor{\bsnm{Jin}, \binits{B.-Q.}},
\bauthor{\bsnm{Korepin}, \binits{V.E.}}:
\batitle{Quantum {S}pin {C}hain, {T}oeplitz {D}eterminants and the
  {F}isher-{H}artwig {C}onjecture}.
\bjtitle{Journal of Statistical Physics}
\bvolume{116},
\bfpage{79}--\blpage{95}
(\byear{2004}).
\doiurl{10.1023/B:JOSS.0000037230.37166.42}
\end{barticle}
\endbibitem

%%% 29
\bibitem{rodney2013}
\begin{barticle}
\bauthor{\bsnm{Rodney}, \binits{M.}},
\bauthor{\bsnm{Song}, \binits{H.F.}},
\bauthor{\bsnm{Lee}, \binits{S.-S.}},
\bauthor{\bsnm{Le~Hur}, \binits{K.}},
\bauthor{\bsnm{S\o{}rensen}, \binits{E.S.}}:
\batitle{Scaling of entanglement entropy across {L}ifshitz transitions}.
\bjtitle{Phys. Rev. B}
\bvolume{87},
\bfpage{115132}
(\byear{2013}).
\doiurl{10.1103/PhysRevB.87.115132}
\end{barticle}
\endbibitem

%%% 30
\bibitem{frahm1992}
\begin{barticle}
\bauthor{\bsnm{Frahm}, \binits{H.}}:
\batitle{Integrable spin-$1/2$ ${XXZ}$ {H}eisenberg chain with competing
  interactions}.
\bjtitle{Journal of Physics A: Mathematical and General}
\bvolume{25}(\bissue{6}),
\bfpage{1417}
(\byear{1992})
\end{barticle}
\endbibitem

%%% 31
\bibitem{grytsiuk}
\begin{barticle}
\bauthor{\bsnm{Grytsiuk}, \binits{S.}},
\bauthor{\bsnm{Hanke}, \binits{J.-P.}},
\bauthor{\bsnm{Hoffmann}, \binits{M.}},
\bauthor{\bsnm{Bouaziz}, \binits{J.}},
\bauthor{\bsnm{Gomonay}, \binits{O.}},
\bauthor{\bsnm{Bihlmayer}, \binits{G.}},
\bauthor{\bsnm{Lounis}, \binits{S.}},
\bauthor{\bsnm{Mokrousov}, \binits{Y.}},
\bauthor{\bsnm{Bl{\"u}gel}, \binits{S.}}:
\batitle{Topological--chiral magnetic interactions driven by emergent orbital
  magnetism}.
\bjtitle{Nature Communications}
\bvolume{11}(\bissue{1}),
\bfpage{1}--\blpage{7}
(\byear{2020})
\end{barticle}
\endbibitem

%%% 32
\bibitem{pachos}
\begin{barticle}
\bauthor{\bsnm{Pachos}, \binits{J.K.}},
\bauthor{\bsnm{Plenio}, \binits{M.B.}}:
\batitle{Three-spin interactions in optical lattices and criticality in cluster
  {H}amiltonians}.
\bjtitle{Physical Review Letters}
\bvolume{93}(\bissue{5}),
\bfpage{056402}
(\byear{2004})
\end{barticle}
\endbibitem

%%% 33
\bibitem{ozawa2019}
\begin{barticle}
\bauthor{\bsnm{Ozawa}, \binits{T.}},
\bauthor{\bsnm{Price}, \binits{H.M.}},
\bauthor{\bsnm{Amo}, \binits{A.}},
\bauthor{\bsnm{Goldman}, \binits{N.}},
\bauthor{\bsnm{Hafezi}, \binits{M.}},
\bauthor{\bsnm{Lu}, \binits{L.}},
\bauthor{\bsnm{Rechtsman}, \binits{M.C.}},
\bauthor{\bsnm{Schuster}, \binits{D.}},
\bauthor{\bsnm{Simon}, \binits{J.}},
\bauthor{\bsnm{Zilberberg}, \binits{O.}}, \betal:
\batitle{Topological photonics}.
\bjtitle{Reviews of Modern Physics}
\bvolume{91}(\bissue{1}),
\bfpage{015006}
(\byear{2019})
\end{barticle}
\endbibitem

%%% 34
\bibitem{mankovsky}
\begin{barticle}
\bauthor{\bsnm{Mankovsky}, \binits{S.}},
\bauthor{\bsnm{Polesya}, \binits{S.}},
\bauthor{\bsnm{Ebert}, \binits{H.}}:
\batitle{Extension of the standard {H}eisenberg {H}amiltonian to multispin
  exchange interactions}.
\bjtitle{Physical Review B}
\bvolume{101}(\bissue{17}),
\bfpage{174401}
(\byear{2020})
\end{barticle}
\endbibitem

\end{thebibliography}

\end{document}